\documentclass[10pt, amssymb, nofootinbib, aps, prd,twocolumn,preprintnumbers]{revtex4-2}
\usepackage{graphicx} 
\usepackage{amsmath} 
\usepackage{xcolor} 
\usepackage{braket} 
\usepackage{tikz} 
\usetikzlibrary{shapes,shapes.geometric,arrows,positioning,automata,backgrounds,calc,er,patterns,snakes,decorations.pathmorphing}
\usepackage{tikz-feynman}

\newcommand{\X}{{\mathbf x}}
\newcommand{\A}{{\mathbf a}}
\newcommand{\K}{{\mathbf k}}
\newcommand{\Ss}{{\mathbf S}}

\newcommand{\V}{{\mathbf v}}

\newcommand{\RR}{{\mathbf n}}
\newcommand{\ds}{\displaystyle}
\newcommand{\vphi}{\varphi}

\usepackage[linktocpage]{hyperref}

\allowdisplaybreaks[3]

\usepackage{pgfplots}
\pgfplotsset{compat=1.16}

\begin{document}

\title{Post-Newtonian Dynamics of Spinning Black Hole Binaries in Einstein-Scalar-Gauss-Bonnet Gravity}

\author{Gabriel Luz Almeida}
\email{galmeida@ustc.edu.cn}	
\affiliation{Interdisciplinary Center for Theoretical Study,
University of Science and Technology of China, Hefei, Anhui 230026, China}
\affiliation{Peng Huanwu Center for Fundamental Theory, Hefei, Anhui 230026, China}

\author{Shuang-Yong Zhou}
\email{zhoushy@ustc.edu.cn}	
\affiliation{Interdisciplinary Center for Theoretical Study,
University of Science and Technology of China, Hefei, Anhui 230026, China}
\affiliation{Peng Huanwu Center for Fundamental Theory, Hefei, Anhui 230026, China}

\begin{abstract}
We explore the post-Newtonian dynamics of spinning black hole (BH) binaries in Einstein-Scalar-Gauss-Bonnet (ESGB) gravity, a theory that modifies general relativity by introducing a massless scalar field coupled nonminimally to gravity via the Gauss-Bonnet term. By employing an effective field theory (EFT) approach, we extend the Routhian formalism to incorporate spin effects in scalar interactions. In this formalism, we derive for the first time the effective two-body potential for spinning BHs up to the third post-Newtonian (3PN) order in generic ESGB gravity theories. This potential is expressed in terms of the sensitivities of the BHs, which are then obtained through a matching procedure using analytic BH solutions derived here within a slow-rotation approximation, accurate to seventh order in the Gauss-Bonnet coupling and fifth order in the BH spin. We also examine the thermodynamic properties of these rotating solutions, which, as shown in previous work, yield important insights into the inspiral phase evolution in ESGB gravity. 
\end{abstract}

\preprint{USTC-ICTS/PCFT-24-26}

\maketitle

\tableofcontents

\section{Introduction}

The detection of gravitational waves (GWs) from binary black hole (BH) mergers \cite{LIGOScientific:2021djp}
by the LIGO-Virgo-KAGRA collaboration \cite{TheLIGOScientific:2014jea,TheVirgo:2014hva,KAGRA:2020tym}
has opened a new era in astrophysics, allowing us to probe the dynamics of these extreme systems with unprecedented precision 
\cite{LIGOScientific:2021sio}.
These observations provide a unique opportunity to test the limit of general relativity (GR) in the strong-field regime, where deviations might appear.

Among the many alternative theories of gravity (see Ref.~\cite{Berti:2015itd} for a comprehensive introduction), Einstein-Scalar-Gauss-Bonnet (ESGB) gravity, which modifies GR by introducing a scalar field $\vphi$ coupled to the Gauss-Bonnet invariant $\mathcal{G} = R^{\mu\nu\rho\sigma} R_{\mu\nu\rho\sigma} - 4 R^{\mu\nu} R_{\mu\nu} + R^2$ in the form of $\alpha f(\vphi) \mathcal{G}$ has attracted considerable interest in recent years. This is mainly because, being some of the lowest order EFT corrections, it allows for BH solutions to present nontrivial scalar field profiles \cite{Sotiriou_2014, Sotiriou_2014b, Kanti:1995vq, Yunes:2011we}, thereby evading the no-hair theorems of GR \cite{Israel:1967wq,Carter:1971zc,
Bekenstein:1995un, Sotiriou:2011dz, Hui:2012qt} (see \cite{Herdeiro:2015waa} for a review) and giving rise to BH spontaneous scalarization \cite{Doneva:2017bvd, Silva:2017uqg} (see \cite{Doneva:2022ewd} for a review). 
Additionally, the coupling leads to second-order field equations \cite{Kobayashi:2011nu,Kobayashi:2019hrl}, making it a subclass of the Horndeski's theory \cite{Horndeski:1974wa}, and  
is also well-motivated as it emerges from the low-energy limit of heterotic string theory \cite{Gross:1986mw,METSAEV1987385}.
All this makes ESGB gravity a compelling theory for exploring deviations from GR, especially in the context of the strong-field phenomena in BH mergers.

To date, the analytic study of BH binaries in ESGB gravity has been exclusively focused on non-rotating, ``static" BHs \cite{BertiJulie2019,Julie:2022qux,Julie:2024fwy}, while for numerical simulations involving rotating BHs only a few specific cases in ESGB gravity have been explored \cite{Okounkova:2020rqw,East:2020hgw,East:2021bqk}. However, in realistic astrophysical scenarios, BHs have significant angular momentum, making a more comprehensive study of spin effects particularly important. In this case, the scalar hair of the rotating BHs will affect the gravitational waveforms emitted by binaries, potentially allowing for signatures of ESGB gravity due to spin to be detected by upcoming GW observatories (e.g., the Einstein Telescope \cite{Punturo:2010zz}  
and LISA \cite{LISA:2022kgy}).

To analyze the dynamics of spinning BH binaries in ESGB gravity, we employ the post-Newtonian (PN) approximation \cite{Blanchet:2013haa}, a method that expands the equations of motion in powers of the characteristic velocity of the system relative to the speed of light. The PN formalism is well-suited for describing the inspiral phase of compact binaries, where the velocities are still relatively small compared to the speed of light, and the gravitational fields are weak enough for perturbative methods to be applicable.
Within the PN formalism, important progress in the understanding of the coalescence of compact binary systems in GR has been attained in the past few decades at very high precision \cite{Faye:2012we,Jaranowski:2013lca,Bini:2013zaa,Damour:2014jta,Jaranowski:2015lha,Damour:2016abl,Bernard:2015njp,Bernard:2016wrg,Foffa:2019rdf,Foffa:2019yfl,Henry:2021cek,Henry:2022ccf,Almeida:2021xwn,Blumlein:2021txe,Almeida:2022jrv,Almeida:2023yia,Blanchet:2023bwj,Blanchet:2023sbv,Amalberti:2024jaa}.
By extending this formalism to include spin effects in the context of ESGB gravity, we can derive the corrections due to both the scalar field and the BH spins.

To tackle these challenges, we adopt an effective field theory (EFT) approach based on the nonrelativistic general relativity (NRGR) framework by Goldberger and Rothstein \cite{Goldberger:2004jt}, which allows for a systematic incorporation of spin and finite-size effects into the PN framework. The inclusion of spins within this framework was first achieved by Porto in Ref.~\cite{Porto:2005ac}, with significant advancements in subsequent works \cite{rothstein2006prl,rothstein2008spin1spin2,rothstein2008spin1spin1,Levi:2008nh,Porto:2010tr,Porto:2010zg,Levi:2010zu,Levi:2011eq,Porto:2012as,Levi:2015msa}. The EFT formalism is particularly powerful because it provides a clear separation between the short- and long-range gravitational interactions, both of which are treated perturbatively in the PN approximation \cite{Foffa:2021pkg}. This approach not only simplifies the calculations but also provides a transparent interpretation of the various contributions to the dynamics.  For comprehensive reviews on this EFT in the context of GR, the reader is referred to Refs.~\cite{Goldberger:2007hy,Foffa:2013qca,Porto:2016pyg,Levi:2018nxp,Sturani:2021ucg}.

The main goal of the present paper is to extend the EFT framework discussed above to the case of spinning BH binaries in ESGB gravity and derive the spin contributions to the conservative dynamics up to the third PN order. This work builds on previous studies that have explored the conservative dynamics of non-spinning compact binaries in scalar-tensor (ST) theories and ESGB gravity. Currently, the 3PN order represent the state-of-the-art in this sector of ESGB gravity \cite{Julie:2022qux}, which includes the 3PN dynamics obtained in Refs.~\cite{TDamour_1992,Mirshekari:2013vb,Bernard:2018hta,Bernard:2018ivi} in ST theories, in addition to an extra ESGB contribution entering at the 3PN order computed in Ref.~\cite{BertiJulie2019}. See also Ref.~\cite{Almeida:2024uph} for a derivation of the 2PN dynamics in ST theories and the ESGB contribution at 3PN order using the EFT method.

To include spin corrections in ESGB gravity, we extend the Routhian formalism used within the NRGR framework, as introduced by Porto in Ref.~\cite{porto2007proceedings}, to account for the presence of a scalar field. It is well known that in gravitational theories with a scalar degree of freedom, the mass of the bodies, treated as point particles in the PN formalism, becomes a function of the scalar field, $m(\vphi)$ \cite{1975ApJEardley}. The identification of this function with the intrinsic parameters of the BH is achieved through a matching procedure, which makes use of the isolated BH solutions.

In the PN framework, the BH's mass function must be expanded in terms of the scalar field $\vphi$, expressed through coefficients known as sensitivities \cite{Damour:1992we}. These sensitivities characterize each BH's response to the slowly varying scalar field generated by its companion \cite{Julie:2022huo} and directly enter the effective two-body potential. Considering this, we follow Ref.~\cite{Maselli:2015tta} to obtain spinning BH solutions in generic ESGB gravity by employing a slow-rotating approximation, which uses as the small parameter of the expansion the spin $\chi = J/m^2$, with $J$ being the angular momentum and $m$ the mass of the BH. We also consider the weak-coupling limit of ESGB gravity, expanding in terms of the small parameter $\epsilon \equiv \alpha f'(\vphi_0)/(4m^2)$ \cite{BertiJulie2019}. Analytical solutions are then obtained up to $\mathcal{O}(\epsilon^7, \chi^5)$ for generic $f(\vphi)$, from which a matching procedure is employed to determine the BH's mass function. This then allow us to derive for the first time the sensitivities of spinning BHs in ESGB gravity. 
Finally, we compute for the first time all spin contributions to the effective two-body potential up to the 3PN order for generic ESGB theories. 

The paper is organized as follows:
In Sec.~\ref{Sec:ESGBgravity}, we provide an overview of ESGB gravity, including a derivation of spinning BH solutions in these theories. 
Sec.~\ref{sec:spinsinESGBgrav} first introduces the basic EFT formalism and details the PN expansion for non-spinning BHs in ESGB gravity. 
In Appendix~\ref{sec:spinGRreview}, we review the inclusion of spins into NRGR using the Routhian formalism. 
Building upon this, Sec.~\ref{sec:spinsinESGBgrav} then extends it to accommodate the scalar field in ESGB gravity. 
In Sec.~\ref{sec:3PNdynamicsESGB}, we compute the spin corrections due to the scalar field and present the final 3PN dynamics for spinning BH binaries in ESGB gravity, 
concluding with a brief analysis of spin effects on the BH sensitivities. Finally, in Sec.~\ref{sec:conclusion}, we discuss the implications of our findings and outline possible directions for future research.

\section{Black holes in ESGB gravity}\label{Sec:ESGBgravity}


ESGB gravity modifies GR by introducing an extra massless scalar degree of freedom that couples non-minimally to gravity via the Gauss-Bonnet invariant. In the presence of matter and using geometrical units ($G = c = 1$), this theory is described by\footnote{In this paper, we adopt the mostly plus signature for the metric where the Minkowski metric is $\eta_{\mu\nu} = {\rm diag}(-1,1,1,1)$.} 
\begin{align}\label{eq:GBaction}
S = S_{\rm EH}[g_{\mu\nu}] + S_{\vphi}[\vphi,g_{\mu\nu}] + S_m[\Psi, \mathcal{A}^2(\vphi)g_{\mu\nu}]\,,
\end{align}
where $S_{\rm EH}$ is the Einstein-Hilbert action, and $S_{\vphi}$ accounts for the coupling of the scalar field with gravity, written in the Einstein frame as
\begin{align}
S_{\rm EH} = &\frac{1}{16\pi} \int d^4x\,\sqrt{-g} R \,, \label{EHaction0} \\
S_{\vphi} = &\frac{1}{16\pi} \int d^4x\,\sqrt{-g} \left[ - 2g^{\mu\nu}\partial_\mu \vphi \partial_\nu \vphi + \alpha f(\vphi) \mathcal{G} \right]\,.  \label{phiaction0}
\end{align}
The third term in \eqref{eq:GBaction} provides an action for matter fields, represented generically by $\Psi$, which are directly coupled to the Jordan frame metric $\tilde{g}_{\mu\nu} = \mathcal{A}^2(\vphi) g_{\mu\nu}$. 

In these expressions, $g = \det g_{\mu\nu}$ is the metric determinant, and $\mathcal{G}$ is the Gauss-Bonnet invariant defined by
\begin{equation}
\mathcal{G} \equiv R^{\mu\nu\rho\sigma} R_{\mu\nu\rho\sigma} - 4 R^{\mu\nu} R_{\mu\nu} + R^2\,,
\end{equation} 
in terms of the Riemann tensor $R_{\mu\nu\rho\sigma}$, Ricci tensor $R_{\mu\nu}$, and Ricci scalar $R$. The coupling constant $\alpha$ has dimensions of $({\rm length})^2$, and $f$ is a dimensionless function of $\vphi$. A specific ESGB theory is determined once the functions $f(\vphi)$ and $\mathcal{A}(\vphi)$ are specified. Notably, ESGB gravity reduces to generic ST theories when $\alpha = 0$ or $f$ is a constant (since $\int d^Dx\sqrt{-g}\mathcal{G}$ is a boundary term in four-dimensional spacetime), and to GR if, in addition, $\mathcal{A}$ and $\vphi$ are constant.

In vacuum, variation of the action \eqref{eq:GBaction} yields the field equations:
\begin{subequations}\label{eq:fieldeqsESGB}
\begin{align}
	G_{\mu\nu} &= T_{\mu\nu}\,,  \label{eq:fieldequationGT} \\
	\Box \vphi &= - \frac14 \alpha f'(\vphi) \mathcal{G}\,, \label{eq:fieldequationscalar}
\end{align}
\end{subequations}
where  $G_{\mu\nu} = R_{\mu\nu}-\frac12 g_{\mu\nu}R$, and $T_{\mu\nu}$ is the effective stress-energy tensor given by
\begin{equation}\label{eq:energymomentumtensor}
	 T_{\mu\nu} = 2 \nabla_\mu \vphi \nabla_\nu \vphi - g_{\mu\nu} (\nabla\vphi)^2 - 4 \alpha P_{\mu\alpha\nu\beta} \nabla^\alpha \nabla^\beta f\,,
\end{equation}
with $\nabla_\mu$ denoting the covariant derivative compatible with $g_{\mu\nu}$ and $\Box \equiv \nabla^\mu \nabla_\mu$. Primes denote derivatives with respect to the scalar field, as in $f'(\vphi) \equiv df(\vphi)/d\vphi$. The new quantity $P_{\mu\nu\rho\sigma}$ is defined by 
\begin{equation}
P_{\mu\nu\rho\sigma} = R_{\mu\nu\rho\sigma} - 2 g_{\mu[\rho} R_{\sigma]\nu} + 2 g_{\nu[\rho} R_{\sigma]\mu} + g_{\mu[\rho} g_{\sigma]\nu}R\,, 
\end{equation}
and is a divergenceless quantity with the same symmetries as the Riemann tensor.

In what follows, we will obtain analytical, asymptotically flat rotating BH solutions to the field equations \eqref{eq:fieldeqsESGB}. To do this, we will use the approach developed by Hartle in Refs.~\cite{Hartle:1967he,Hartle:1968si}, and further extended by Maselli et al. in Ref.~\cite{Maselli:2015tta}, which generates rotating solutions through a slow-rotation expansion in $\chi \equiv J/m^2$, where $J$ is the angular momentum and $m$ the Arnowitt-Deser-Misner (ADM) mass of the BH, starting from a static background solution. Besides the expansion in $\chi$, we will solve the field equations perturbatively in $\alpha$, which always appears through the dimensionless combination 
\begin{equation}\label{eq:epsilonparameter}
\epsilon \equiv \frac{\alpha f'(\vphi_0)}{4m^2} \ll 1\,,
\end{equation}
where $\vphi_0$ is the asymptotic value of the scalar field. Solutions are then obtained up to $\mathcal{O}(\epsilon^7,\chi^5)$, generalizing the results of Ref.~\cite{Maselli:2015tta}, which provided solutions for the specific case of Einstein-dilaton-Gauss-Bonnet (EDGB) gravity ($f(\vphi) = \frac14 e^{2\vphi}$) at this same order,  $\mathcal{O}(\epsilon^7,\chi^5)$, but now for an unspecified $f(\vphi)$. We begin by deriving analytical static solutions up to $\mathcal{O}(\epsilon^7)$ below.  

Finally, the smallness of $\epsilon$ in Eq.~\eqref{eq:epsilonparameter} is justified by constraints placed by theoretical considerations such as causality bounds \cite{Hong:2023zgm, Tolley:2020gtv, Caron-Huot:2020cmc} and regularity conditions on the BH's event horizon \cite{Kanti:1995vq,Sotiriou_2014b}, as well as from x-ray and GW observations \cite{Yagi:2012gp,Nair:2019iur,Perkins:2021mhb,Wang:2021jfc,Julie:2024fwy}. Within this perturbative approach, BH solutions in ESGB gravity have been extensively studied in both static and slowly-rotating cases, but limited to a few specific theories. Notably, the astrophysical properties of these solutions have been explored analytically up to $\mathcal{O}(\epsilon^7,\chi^5)$  in EDGB gravity \cite{Yunes:2011we,Ayzenberg:2014aka,Pani:2011gy,Maselli:2015tta}, but only up to 
$\mathcal{O}(\epsilon^2,\chi^1)$ \cite{Sotiriou_2014b} in the ``shift symmetric" theory with $f(\vphi) = 2\vphi$.

\subsection{Static BH solutions}\label{subsec:statbhsols}

To obtain static BH solutions, we begin by substituting an ansatz for spherically symmetric spacetimes into the field equations, choosing here the form 
\begin{equation}\label{eq:metricdroste}
ds^2 = - n(r) \sigma^2(r) dt^2 + n^{-1}(r) dr^2 + r^2 d\Omega^2 \,,
\end{equation}
where $d\Omega^2 = d\theta^2+\sin^2\theta d\phi$, with scalar field $\vphi(r)$ depending only on the radial coordinate $r$.

At lowest order in $\epsilon$, the solution is simply that of Schwarzschild. 
The field equations can be solved order by order in $\epsilon$, using expansions for the metric and scalar field truncated at $\mathcal{O}(\epsilon^7)$:
\begin{subequations}\label{eq:metricfunctionsstatic}
\begin{align}
n(r) &= 1 - \frac{2m}{r} + \sum_{i=1}^7 \epsilon^i n_i(r)\,, \label{eq:metricnofr} \\
\sigma(r) &= 1 + \sum_{i=1}^7 \epsilon^i \sigma_i(r)\,, \\
\vphi(r) &= \vphi_0 + \sum_{i=1}^7 \epsilon^i \vphi_i(r)\,.
\end{align}
\end{subequations}
Additionally, we expand the function $f(\vphi)$ around $\vphi_0$ as
\begin{equation}\label{eq:fvphiexpansion}
f(\vphi) = \sum_{i=0}^7 \frac{1}{n!} f^{(n)}(\vphi_0) (\vphi-\vphi_0)^n\,.
\end{equation}
Then, static solutions are obtained by plugging the above expanded metric functions $n(r)$, $\sigma(r)$, $\vphi(r)$, as well as $f(\vphi)$, into the metric \eqref{eq:metricdroste}. This reduces the field equations \eqref{eq:fieldeqsESGB} to a set of second-order ordinary differential equations (ODEs) in $r$, which can be solved iteratively for each of the functions $n_i$, $\sigma_i$, and $\vphi_i$, order by order in $\epsilon$. The solutions are uniquely determined by imposing asymptotically flatness ($g_{\mu\nu}\rightarrow \eta_{\mu\nu}$ and $\vphi\rightarrow \vphi_0$ at spatial infinity) and ensuring the regularity of the scalar field at the BH's horizon.

From the asymptotic behavior of the solutions, we can extract the ADM mass and the scalar charge of the BH. These are defined, respectively, as the coefficients of the $\mathcal{O}(1/r)$ terms in $g_{tt}$ and $\vphi(r)$: 
\begin{subequations}
\begin{align}
	g_{tt} = -1 + \frac{2m}{r} + \mathcal{O}\left( \frac{1}{r^2} \right) \,,  \label{eq:asymbehaviorgtt}\\
	\vphi(r) = \vphi_0 + \frac{D}{r} + \mathcal{O}\left( \frac{1}{r^2} \right) \,. \label{eq:asymbehaviorscalar}
\end{align} 
\end{subequations}
As it turns out, the scalar charge $D$ is not an independent parameter but is entirely determined in terms of the ADM mass $m$ of the BH and the asymptotic scalar field value $\vphi_0$. Despite this, the scalar field profile is nontrivial, resulting in solutions that differ from those in GR. In this case, although there are no new independent charges, the BH is said to be ``hairy," with the scalar charge being a ``secondary type" hair. 

It is worth noting that many BH solutions in the ESGB literature set the asymptotic scalar field value to zero. However, this approach is not applicable in the present context, as emphasized by the authors of Ref.~\cite{BertiJulie2019}, since it will later represent the asymptotic scalar field generated by the companion in a binary system.

Analytical solutions for spherically symmetric BHs in generic ESGB gravity were first obtained in Ref.~\cite{BertiJulie2019} up to $\mathcal{O}(\epsilon^4)$, and have been recently extended to higher orders in Ref.~\cite{Julie:2023ncq}. Our solutions, provided in the ancillary file, match the latter up to $\mathcal{O}(\epsilon^7)$.

\subsection{Rotating BH solutions}\label{subsec:rotBHsols}

We now address the inclusion of rotation for BH solutions in ESGB gravity, following the generic approach presented in Ref.~\cite{Maselli:2015tta}.
In this framework, stationary, axially symmetric spacetimes are constructed via a perturbative expansion around the static solution, using the spin $\chi$ as the small parameter of the expansion. 

Adapted to the coordinates of Eq.~\eqref{eq:metricdroste}, the most general ansatz for such spacetimes - assuming equatorial symmetry and invariance under $(t,\phi) \rightarrow - (t,\phi)$ - is given by
\begin{align}\label{eq:metricrotation}
ds^2 &= - n \sigma^2 [ 1+2 h(r,\theta)] dt^2 + n^{-1} [ 1 + 2M(r,\theta)]  dr^2 \nonumber\\
& + r^2 [ 1+ 2k(r,\theta)] [d\theta^2 + \sin^2\theta \left( d\phi - \hat{\omega}(r,\theta) dt \right)^2]\,.
\end{align}
The new functions of $(r,\theta)$ appearing in this metric are then expanded in terms of a complete basis of orthogonal polynomials, which, given the assumed symmetries, can be written as
\begin{subequations}\label{eq:metricfunctionsrotexpansion}
\begin{align}
\hat{\omega}(r,\theta) &= \sum^{N_{\chi}-q}_{n=1,3,5,\dots} \sum_{l=1,3,5,\dots}^n \chi^n \omega^{(n)}_l(r) S_l(\theta)\,, \\
h(r,\theta) &= \sum^{N_{\chi}-p}_{n=2,4,\dots} \sum_{l=0,2,4,\dots}^n \chi^n h^{(n)}_l(r) P_l(\cos\theta)\,, \\
M(r,\theta) &= \sum^{N_{\chi}-p}_{n=2,4,\dots} \sum_{l=0,2,4,\dots}^n \chi^n M^{(n)}_l(r) P_l(\cos\theta)\,, \\
k(r,\theta) &= \sum^{N_{\chi}-p}_{n=2,4,\dots} \sum_{l=0,2,4,\dots}^n \chi^n k^{(n)}_l(r) P_l(\cos\theta)\,. \label{eq:kofrexpansion}
\end{align}
\end{subequations}
For the scalar field, we similarly have
\begin{equation}\label{eq:scalarfieldwspin}
\vphi(r,\theta) = \bar{\vphi}(r) + \!\! \sum^{N_{\chi}-p}_{n=2,4,\dots} \sum_{l=0,2,4,\dots}^n \!\chi^n \vphi^{(n)}_l(r) P_l(\cos\theta)\,.
\end{equation}
In the above expressions, $P_l$ are the Legendre polynomials, $S_l \equiv -\frac{1}{\sin\theta}\frac{dP_l(\cos\theta)}{d\theta}$, $N_{\chi}$ is the order of the spin expansion, and $p$ and $q$ are numbers taking the values of either $(p,q)=(0,1)$ if $N_{\chi}$ is even or $(p,q)= (1,0)$ if $N_{\chi}$ is odd. In our case, we have $N_{\chi} = 5$, corresponding to $\mathcal{O}(\chi^5)$ and hence $N_{\chi}-p = 4$ and $N_{\chi}-q = 5$.  In Eq.~\eqref{eq:scalarfieldwspin}, the function $\bar{\vphi}(r)$ represents the scalar field obtained in the static case. 

Finally, since the static solution, being perturbative in the Gauss-Bonnet coupling, serves as the seed for spin corrections, the functions $\omega^{(n)}_l$, $h^{(n)}_l$, $M^{(n)}_l$, $k^{(n)}_l$, and $\vphi^{(n)}_l$ must also be expanded in $\epsilon$. Thus, for instance, we have $\omega^{(n)}_l = \sum_{m=0}^7 \epsilon^m \omega^{(n)}_{lm}$, and similarly for the other functions. Note that the functions $k^{(n)}_0$ can be set to zero without loss of generality, as this simply corresponds to a redefinition of the coordinate $r$ in \eqref{eq:metricrotation}.

\subsubsection{$\mathcal{O}(\chi^n)$ corrections}

As in the static case, solutions to the field equations are obtained by substituing the metric ansatz \eqref{eq:metricrotation}, with functions $\hat{\omega}$, $h$, $M$, $k$, and $\vphi$ expanded according to Eqs.~\eqref{eq:metricfunctionsrotexpansion} and \eqref{eq:scalarfieldwspin}, into Eq.~\eqref{eq:fieldeqsESGB}, and solving the resulting equations. This results in a set of second-order differential equations for the functions $\omega^{(n)}_{lm}$, $h^{(n)}_{lm}$, $M^{(n)}_{lm}$, $k^{(n)}_{lm}$, and $\vphi^{(n)}_{lm}$ for all orders $(\chi^n,\epsilon^m)$ of our expansion. Thus, for every choice of parameters $(m,n)$, the set of differential equations will depend on the lower-order solutions, as well as on the functions $n_i$, $\sigma_i$, and $\vphi_i$ (with $i\le m$) of the spherically symmetric case.

For the $\mathcal{O}(\chi^1)$ correction, the metric is given by 
\begin{equation}
ds^2 = ds^2_{\rm static} - 2 r^2 \omega(r) \sin^2\theta dt d\phi\,,
\end{equation}
where $ds^2_{\rm static}$ is the metric of the static case \eqref{eq:metricdroste} and $\omega(r) \equiv \omega^{(1)}_1$ is the gravitomagnetic term. The solution for this term can be easily obtained at every order of $\epsilon$ by solving the $(t\phi)$ component of the field equation \eqref{eq:fieldequationGT}. From this solution, the angular momentum of the BH can be determined from the asymptotic behavior of $\omega(r)$ at large distances:
\begin{equation}\label{eq:asymbehavioromega}
\omega(r) \rightarrow \frac{2J}{r^3}\,.
\end{equation}

For the $\mathcal{O}(\chi^n)$ corrections with even $n\ge 2$, we project the field equations onto the Legendre polynomails $P_l(\cos\theta)$, for $l = 0, 2,\dots, n$, as follows:
\begin{equation}
\int_0^\pi d\theta\, \sin\theta P_l(\cos\theta)\left(G_{\mu\nu}-T_{\mu\nu}\right) = 0\,,
\end{equation}
and the same for the scalar field equation.
By doing this, our set of differential equations in $(r,\theta)$ reduces to a set of ODEs in the radial coordinate $r$ only, for the functions $h^{(n)}_{lm}(r), M^{(n)}_{lm}(r), k^{(n)}_{lm}(r)$, and $\vphi^{(n)}_{lm}(r)$. In this case, besides the scalar field equation \eqref{eq:fieldequationscalar}, it suffices to consider only the diagonal components of the field equations \eqref{eq:fieldequationGT}. For corrections with odd $n\ge 3$, on the other hand, the needed ODEs, for the functions $\omega^{(n)}_{lm}(r)$, are obtained by projecting the $(t\phi)$ component of \eqref{eq:fieldequationGT} onto $S_l(\cos\theta)$, for $l = 1, 3,\dots, n$:
\begin{equation}
\int_0^\pi d\theta\, \sin\theta S_l(\cos\theta)\left(G_{t\phi}-T_{t\phi}\right) = 0\,.
\end{equation}

The procedure outlined above yields analytical solutions for all functions in the expansion of the metric and scalar field. These solutions are unique, provided we impose the following conditions: (i) the spacetime metric is asymptotically flat, and the scalar field takes the background value $\vphi_0$ at spatial infinity; (ii) the scalar field is regular at the BH's event horizon; and (iii) integration constants arising from solving the ODEs are absorbed into the definitions of the ADM mass $m$, angular momentum $J$, and scalar background value $\vphi_0$, according to the $1/r$ behavior of the metric and scalar field shown in Eqs.~\eqref{eq:asymbehaviorgtt}, \eqref{eq:asymbehaviorscalar}, and \eqref{eq:asymbehavioromega}.

The solutions derived here are accurate up to $\mathcal{O}(\epsilon^7,\chi^5)$ and generalize the results of Ref.~\cite{Maselli:2015tta}, where expressions for the specific case of Einstein-dilaton-Gauss-Bonnet were obtained, to the most generic ESGB gravity. These solutions depend only on the constants $m$, $J$, and $\vphi_0$, as well as on the derivatives $f^{(n)}(\vphi_0)$. Due to the sizes of these expressions, we have made them available in a Mathematica notebook in the supplemental material.

\subsection{Properties of the solutions}

As we will see in Sec.~\ref{subsec:ESGBPNapprox}, in the PN description of compact objects in a binary system, each object is effectively treated as a point particle. From the EFT point of view, this entails describing them by operators located on the bodies' worldlines, with internal physics encoded in the Wilson coefficients. The connection between the point-particle description and the BH solutions is provided by a matching procedure, which is necessary to determine the Wilson coefficients in terms of the internal properties of each of the compact objects. For this reason, before moving to the EFT description, we derive both geometrical and thermodynamical quantities of the BHs in ESGB gravity, which will be crucial for understanding the physics behind the matching.

\subsubsection{Geometrical properties}\label{subsubssec:geomproperties}

We start by considering the event horizon radius $r_{\rm h}$, defined as the largest root of $g_{\phi\phi} g_{tt}- g_{t\phi}^2 = 0$. The computation, up to $\mathcal{O}(\epsilon^7,\chi^5)$, results in
\begin{equation}\label{eq:horizoncoeffs}
r_{\rm h} = 2m \left( 1 + \sum_{n=1}^7 \sum_{l = 0,2,4} \epsilon^n \chi^l r^{(n)}_{l} \right)\,,
\end{equation}
where $r^{(n)}_{l}$ depend on the specific ESGB theory only through derivatives of $f(\vphi)$ evaluated at $\vphi_0$, denoted here generically by $f^{(k)}(\vphi_0)$. 

Next, we derive the angular velocity of the BH horizon, obtained via
\begin{equation}
	\Omega_{\rm h} = - \lim_{r\rightarrow r_{\rm h}} \frac{g_{t\phi}}{g_{\phi\phi}}\,,
\end{equation}
from which we can compute the BH's moment of inertia, $I = J/\Omega_{\rm h}$. The results have general form given by 
\begin{align}
\Omega_{\rm h} &= \frac{\chi}{4m} \left( 1 + \sum_{n=1}^7 \sum_{l = 0,2} \epsilon^n \chi^l \Omega^{(n)}_{l} \right)\,, \\
I &= 4 m^3 \left( 1 + \sum_{n=1}^7 \sum_{l = 0,2} \epsilon^n \chi^l I^{(n)}_{l} \right)\,.
\end{align}
As in Eq.~\eqref{eq:horizoncoeffs}, the coefficients $\Omega^{(n)}_{l}$ and $I^{(n)}_{l}$ depend on the theory through $f^{(k)}(\vphi_0)$. The results for all these quantities, as well as those defined below, valid for up to $\mathcal{O}(\epsilon^7,\chi^5)$, are lengthy and, because of this, are only presented in the ancillary file.  

The scalar charge $D$ of the BH can be obtained from the asymptotic behavior of the scalar field  $\vphi(r,\theta)$ [cf. Eq.~\eqref{eq:asymbehaviorscalar}], and has a general form given by
\begin{equation}\label{eq:scalarchargeDsol}
D = 2m \left( \sum_{n=1}^7 \sum_{l = 0,2,4} \epsilon^n \chi^l D^{(n)}_{l} \right)\,,
\end{equation}
where, again, the coefficients $D^{(n)}_{l}$ depend on the particular ESGB theory. From this expression, we immediatelly notice that, for a given ESGB theory, the scalar charge is completely determined by the values of the ADM mass $m$ (entering not only through the multiplicative factor of $2m$ but also in $\epsilon$, which, we recall, is given by $\epsilon = \alpha f'(\vphi_0)/(4m^2)$) and the background scalar field  value $\vphi_0$. This justifies classifying these hairy solutions as being of secondary type.

Finally, we compute the quadrupole moments of the spacetime, induced by rotation in both gravity and scalar sectors, following Ref.~\cite{Maselli:2015tta}. The authors there use a general approach described in Ref.~\cite{Thorne:1980ru} to derive the quadrupole moments from the asymptotic behavior of the metric. In this approach, a preferred coordinate transformation must be performed so as to ensure that the metric components $g_{tt}$ and $g_{ij}$ (with $i,j\neq t$) do not depend on the angular coordinates up to $\mathcal{O}(1/r^2)$ terms. Remarkably, the coordinate transformation found in Ref.~\cite{Maselli:2015tta} for the dilaton case does not involve the Gauss-Bonnet coupling $\alpha$. Because of this, the same transformation can be used in our most general case of ESGB gravity:
\begin{align*}
r &\rightarrow r + \frac{\chi^2 m^2}{2r} \left[ 1 + \frac{m}{r} - \frac{2m^2}{r^2} + \frac{m(6m-r)}{r^2} \cos^2\theta \right]\,, \\
\theta &\rightarrow \theta + \frac{\chi^2 m^3}{r^3} \sin\theta \cos\theta\,.
\end{align*}
Indeed, it is easy to check that such transformation satisfies the requirements of Ref.~\cite{Thorne:1980ru} for the appropriate coordinate choice. The metric $Q^{20}_{g}$ and scalar $Q^{20}_{\vphi}$ quadrupole moments are then derived from the components
\begin{subequations}
\begin{align}
	g_{tt} &= -1 + \frac{2m}{r} + \frac{\sqrt{3}}{2r^3} \left[ Q^{20}_{g} Y^{20} + (l = 0_{\rm pole}) \right] \,, \label{eq:gttquad} \\
	\vphi & = \vphi_0 + \frac{\sqrt{3}}{2r^3} \left[ Q^{20}_{\vphi} Y^{20} + (l = 0_{\rm pole}) \right]  \,, \label{eq:vphiquad}
\end{align}
\end{subequations}
which are accurate to $\mathcal{O}(m^3/r^3)$, and $Y^{20}$ is the $(l,m)=(2,0)$ spherical harmonic. This yields 
\begin{subequations}
\begin{align}
	Q^{20}_{g} &= - \sqrt{\frac{64\pi}{15}} \chi^2 m^3  \left( 1 + \sum_{n=1}^7 \sum_{l = 0,2} \epsilon^n \chi^l Q^{(n)}_{g,l} \right)\,,  \label{eq:quadrupolegrav}\\	
	Q^{20}_{\vphi} &= - \sqrt{\frac{64\pi}{15}} \chi^2 m^3  \left( \sum_{n=1}^7 \sum_{l = 0,2} \epsilon^n \chi^l Q^{(n)}_{\vphi,l} \right) \label{eq:quadrupolescalar} \,,
\end{align}
\end{subequations}
where, again, $Q^{(n)}_{g,l}$ and $Q^{(n)}_{\vphi,l}$ are real coefficients. These results are available in the ancillary file. 

As we will see later, the expressions for the two quadrupole moments in Eqs.~\eqref{eq:quadrupolegrav} and \eqref{eq:quadrupolescalar} will be relevant for matching the BH solutions to the finite-size operators that will be part of the point-particle description in the EFT framework.

The results obtained in this section can be directly compared with those presented in Ref.~\cite{Maselli:2015tta} for the EDGB gravity by setting $f(\vphi) = \frac{e^{2\vphi}}{4}$ and identifying their scalar field $\Phi$ with ours through $\Phi = 2\vphi$. In particular, we are in agreement with their expressions for the event horizon and dilaton charge, as presented in Table II of Ref.~\cite{Maselli:2015tta}, as well as the moment of inertia and quadrupole moment respectively in Eqs.~(41) and (42) of their paper.\footnote{However, these last two quantities contain typos in Ref.~\cite{Maselli:2015tta}: for the moment of inertia in Eq.~(41) of Ref.~\cite{Maselli:2015tta}, the sign in front of the $\chi^4$ term should be a minus sign, while in Eq.~(43), for the quadrupole moment, the number 0.1061619 should be replaced with 0.1062619. This can be verified using the ancillary files provided in the supplemental material of Ref.~\cite{Maselli:2015tta}.}

\subsubsection{Thermodynamical properties}

Now, we study the thermodynamical properties of the rotating BH solutions found above for generic ESGB gravity theories. As it was shown in Ref.~\cite{BertiJulie2019}, non-rotating BH solutions in such theories satisfy the first law of BH thermodynamics, provided that the Wald entropy \cite{Wald:1993nt} be considered, which is the appropriate definition for a BH entropy in this case. In particular, here we extend these results to accomodate rotation.

For stationary and axisymmetric spacetimes, thermodynamical quantities are defined in terms of the two natural Killing vectors $\xi^\mu = (\partial/\partial t)^\mu$ and $\psi^\mu = (\partial/\partial \phi)^\mu$, as well as $\chi^\mu = (\partial/\partial t)^\mu + \Omega_{\rm h} (\partial/\partial \phi)^\mu$, which is also a Killing vector and generator of the null geodesics of the BH's event horizon. From this, we define the surface gravity $\kappa$ of the BH as
\begin{equation}
\kappa^2 \equiv \left. - \frac12 (\nabla_\mu \chi_\nu) (\nabla^\mu \chi^\nu) \right|_{r_{\rm h}},
\end{equation}
which allows us to define the BH temperature via
\begin{equation}
T \equiv \frac{\kappa}{2\pi}\,.
\end{equation}
Direct computation of this quantity yields an expression that does not depend on the angular variables, showing that it is indeed an appropriate definition for the temperature of rotating BHs in ESGB gravity.

The Wald entropy $S_{\rm w}$ of a BH in generic ESGB gravity can be written as \cite{Maeda:2009uy,Kleihaus:2015aje}
\begin{equation}
S_{\rm w} = \frac14 \int_{r_{\rm h}} d\Sigma_{\rm h} \left[ 1 + 2\alpha f(\vphi) \tilde{R} \right] \,,
\end{equation} 
where $d\Sigma_{\rm h}$ is the line element of the induced metric on the BH's event horizon, given by
\begin{equation}
d\Sigma_{\rm h} = r^2_{\rm h} \left[ 1 + 2 k(r_{\rm h},\theta) \right] \left( d\theta^2 + \sin^2\theta d\phi^2 \right)
\end{equation}
in our case, and $\tilde{R}$ is the associated 2D Ricci scalar. It is easy to see from this that the Wald entropy is equal to Bekenstein's, $S_{\rm bh} = A/4$, being $A$ the area of the event horizon, plus a term depending on the scalar field $\vphi$, proportional to $\alpha$. In particular, by setting the function $k(r_{\rm h}, \theta)$ to zero in the expression above, (and using $\tilde{R}=2/r_{\rm h}^2$ for the sphere) we readly recover the the Wald entropy for non-rotating BHs in ESGB gravity obtained in Ref.~\cite{BertiJulie2019}:
\begin{equation}
S_{\rm w} = S_{\rm bh} + 4\pi \alpha f(\vphi_{\rm h})\,,
\end{equation}
where $\vphi_{\rm h}\equiv \vphi(r_{\rm h})$.

Once all the above quantities have been computed, the first law of BH thermodynamics can be checked. In particular, it is well known that, for gravity theories which has an additional scalar field, the first law of BH thermodynamics is generized to \cite{Gibbons:1996af} 
\begin{equation}
dm = T dS_{\rm w} + \Omega_{\rm h} dJ - D d\vphi_0\,,
\end{equation}
which contains a term depending on the scalar charge $D$, and where, as mentioned above, the entropy in question should be the Wald's one $S_{\rm w}$. Indeed, we have derived all the necessary quantities using the formulas above, with results available in the anciliary file, and explicitly verified the validity of the first law by computing variations with respect to the ADM mass $m$, the angular momentum $J$, and scalar field environment $\vphi_0$.
This check, performed to $\mathcal{O}(\epsilon^7,\chi^5)$ accuracy, provides a strong self-consistency test for the rotating BH solutions obtained in this section.

As we will see in the following sections, the matching procedure allows us to describe the mass functions of BHs, treated as point particles in the PN approximation, through intrinsic quantities. These quantities are directly related to the thermodynamic properties discussed above, with the first law of BH thermodynamics serving as the principal guide. In the next section, we introduce the BH description in the PN approximation, followed by an EFT approach to model the dynamics of binary systems in ESGB gravity.


\section{EFT framework for spinning BH binaries}\label{sec:spinsinESGBgrav}
	
\subsection{ESGB gravity in PN approximation}\label{subsec:ESGBPNapprox}

In the PN formalism, spinless compact bodies in a binary system are initially modeled as pointlike objects, with spin, finite size, and other corrections added subsequently. Neglecting for now these additional effects, when scalar field couplings are incorporated into this framework, each particle’s mass $m_{\rm a}(\vphi)$, with ${\rm a} = 1,2$ labeling the two bodies, becomes dependent on the scalar field $\vphi$ evaluated on the worldline $x^\mu_{\rm a}(\lambda_{\rm a})$, following from the skeletonization procedure \cite{1975ApJEardley}. This mass dependency reflects the internal structure of the compact objects \cite{BertiJulie2019}, and the matter coupling is expressed via the point-particle action\footnote{Note that the function $\mathcal{A}(\varphi)$ connecting the Einstein and Jordan frames is automatically considered in the definition of the mass in Eq.~\eqref{eq:sourceaction}, and only when translating quantities from one frame to the other is that we need to invoke the explicit definition of of this function; See Ref.~\cite{Julie:2022qux} for more details. }: 
\begin{equation}\label{eq:sourceaction} 
S^{\rm pp}_m[g_{\mu\nu},\vphi,\{x_{\rm a}^\mu\}] = - \sum_{\rm a} \int d\lambda_{\rm a}\,  m_{\rm a}(\vphi)  \sqrt{-u_{\rm a}^2} \,.
\end{equation} 
In this expression, $\lambda_{\rm a}$ is an arbitrary parametrization for the worldline of particle $\rm a$ and $u^{\mu}_{\rm a}=dx^{\mu}_{\rm a}/d\lambda_{\rm a}$.   
Then, the perturbative character of the PN approximation motivates us to define the so-called sensitivity parameters $\alpha_{\rm a}^0, \beta_{\rm a}^0, \beta_{\rm a}'^0, \dots$, which are responsible for measuring the coupling strength of each body to the scalar field and are defined through the functions 
\begin{subequations}
\begin{align}
\alpha_{\rm a} &\equiv \frac{d\ln m_{\rm a}(\vphi)}{d\vphi}\,,  \\
\beta_{\rm a} &\equiv \frac{d\alpha_{\rm a}(\vphi)}{d\vphi}\,,  \\
\beta_{\rm a}' &\equiv \frac{d\beta_{\rm a}(\vphi)}{d\vphi}\,, 
\end{align}
\end{subequations}
and so on, by taking their values at the scalar field background $\vphi_0$:
$\alpha_{\rm a}^0 = \alpha_{\rm a}(\vphi_0)$, $\beta_{\rm a}^0 = \beta_{\rm a}(\vphi_0)$, and similarly for higher-order terms. Hence, the Taylor expansion of $m_{\rm a}(\vphi)$ assumes the form
\begin{align}
m_{\rm a}(\vphi) &= m^0_{\rm a} \bigg[ 1 + \alpha_A^0 (\vphi - \vphi_0) \\
&+ \frac12 \left((\alpha_A^0)^2 + \beta_A^0\right)(\vphi - \vphi_0)^2 \bigg] + \mathcal{O}((\vphi - \vphi_0)^3)\,. \nonumber
\end{align}

It is well-known that the conservative dynamics of compact binaries in ST theories coincides with that of the ESGB gravity up to the 3PN order, when just an extra term must be added to the two-body Lagrangian. In particular, completion of the 3PN dynamics in ESGB gravity for non-spinning bodies was achieved in Ref.~\cite{Julie:2022qux}, building on the ST results at 1PN order by Damour and Esposito-Farèse \cite{TDamour_1992}, 2PN order by Mirshekari and Will \cite{Mirshekari:2013vb}, and 3PN order by Bernard \cite{Bernard:2018hta,Bernard:2018ivi}.
Note that all these results were obtained in the Jordan frame, necessitating careful translation to compare with the Einstein-frame results.

Before reviewing the NRGR framework adapted to the ESGB gravity \cite{Almeida:2024uph}, we introduce some essential notation: $\X_{\rm a}$ denotes the position of body ${\rm a}$, $\V_{\rm a} = \dot{\X}_{\rm a} = d\X_{\rm a}/dt$ the velocity, and $\A_{\rm a} = \dot{\V}_{\rm a}$ the acceleration. The orbital separation between the two objects is denoted by $r = |\X_1-\X_2|$ and $\RR = (\X_1 - \X_2)/r$, and the relative velocity is denoted by $\V = \V_1-\V_2$ and $v = |\V|$.

\subsection{Basic EFT setup}\label{sec:EFTsetup}

The non-relativistic general relativity (NRGR) EFT developed by Goldberger and Rothstein in \cite{Goldberger:2004jt} has been extended to ST theories by several authors \cite{Kuntz:2019zef,Diedrichs:2023foj,Bhattacharyya:2023kbh,Bernard:2023eul} and more recently to the ESGB case in Ref.~\cite{Almeida:2024uph}. Here, we review the fundamental elements presented in the latter reference, which are necessary for extending the theory to include spin. 

Within the PN approximation, perturbative calculations are performed by considering metric fluctuations $h_{\mu\nu}$ around Minkowski spacetime $\eta_{\mu\nu}$ and scalar field fluctuations $\delta\vphi$ around a background value $\vphi_0$:
\begin{equation}\label{fluctuations}
g_{\mu\nu} = \eta_{\mu\nu} + h_{\mu\nu}\,, \qquad \vphi = \vphi_0 + \delta\vphi\,.
\end{equation} 
The fluctuations in Eq.~\eqref{fluctuations} can be split into potential and radiation parts using the method of regions \cite{beneke1998,smirnov2002}, which plays a central role in the EFT formulation:
\begin{equation}\label{methregionsplitting}
h_{\mu\nu} = H_{\mu\nu} + \bar{h}_{\mu\nu} \qquad \text{and} \qquad \delta\vphi = \Phi + \bar{\vphi}\,.
\end{equation}
The potential modes $H_{\mu\nu}$ and $ \Phi$ are defined in the off-shell momentum region of $k^\mu = (k^0,\K) \sim (v/r, 1/r)$ and are responsible for controlling the binding forces that maintain the two objects together. The radiation modes $\bar{h}_{\mu\nu}$ and $\bar{\vphi}$, on the other hand, live in the on-shell region characterized by $k^\mu  \sim (v/r, v/r)$ and represent radiation propagating to infinity. The use of the method of regions is justified by the fact that the inspiraling dynamics of compact binaries present the clear separation of scales  $r_s \ll r \ll \lambda$, valid in the nonrelativistic limit of $v\ll 1$, where $r_s$ is the typical size of the compact object and $\lambda$ is the wavelength of emitted gravitational radiation. This separation of scales then allows us to describe the dynamics of different momentum regions separately, each of them scaling homogeneously with $v$, the natural small parameter of the PN approximation.

The dynamics of a compact binary system is then described by the point-particle action \eqref{eq:sourceaction}, with gravitational and scalar interactions between the bodies ${\rm a} = 1,2$, as well as radiation emitted by the system, governed by $S_{\rm EH}$ and $S_{\vphi}$ in Eqs.~\eqref{EHaction0} and \eqref{phiaction0}. In particular, the gravitational sector must be augmented by a gauge-fixing term, needed to fix the freedom in the choice of coordinates. Here we choose to work in the harmonic gauge:
\begin{equation}
S_{\rm EH+GF} \equiv \int \frac{d^4x\,\sqrt{-g}}{16\pi G} \left( R - \frac{1}{2} \Gamma^\mu \Gamma^\nu g_{\mu\nu} \right)\,,
\end{equation}
where $\Gamma^\mu$ is defined in terms of the Christoffel symbol by $\Gamma^\mu \equiv \Gamma^\mu{}_{\alpha\beta} g^{\alpha\beta}$.

Following \cite{Almeida:2024uph}, we work in $d+1$ spacetime dimensions and employ dimensional regularization. For the metric perturbation, we use the decomposition of \cite{Kol:2007bc}, in which $g_{\mu\nu}$ is parametrized via a scalar $\phi$, a spatial vector $A_i$, and a symmetric spatial tensor $\sigma_{ij}$ by
\begin{eqnarray}\label{kkdecomposition}
g_{\mu\nu} = e^{\frac{2\phi}{\Lambda}}
\left(
\begin{array}{cc}
  -1 & \dfrac{A_i}\Lambda\\
  \dfrac{A_j}\Lambda &  e^{-\frac{c_d\phi}{\Lambda}}\left(\delta_{ij}+\dfrac{\sigma_{ij}}\Lambda\right) - \dfrac{A_iA_j}{\Lambda^2}
\end{array}
\right) \,,
\end{eqnarray}
with fields normalized by $\Lambda \equiv (32\pi G)^{-1/2}$ and $c_d \equiv 2(d-1)/(d-2)$.  This parameterization is known to improve efficiency in some cases by reducing the number of required Feynman diagrams (See, e.g., \cite{Gilmore:2008gq,Foffa:2011ub}).  Likewise, it is also convenient to rescale  $\delta\vphi \rightarrow \delta\vphi/\Lambda$.

Since here we are interested only in the conservative sector, which is described by the potential region, we can judiciously set to zero the radiation modes. Hence, corrections to the two-body effective action $S_{\rm eff}$ are obtained by integrating out the potential modes, 
\begin{align}
e^{i S_{\rm eff}[x_{\rm a}^\mu]} = &\int \mathcal{D}\Phi\mathcal{D}\phi \mathcal{D}A_i \mathcal{D}\sigma_{ij} \exp\big[ i ( S_{\rm EH+GF}[g_{\mu\nu}] \nonumber\\ 
&+ S_{\vphi}[g_{\mu\nu},\Phi] + S_m[g_{\mu\nu},\Phi,{x_{\rm a}^\mu}] ) \big] \,. \label{pathintegralforSTtheories}
\end{align}
From this path integral, Feynman rules for propagators, field self-interactions, and worldline couplings can be derived and organized in powers of $v$. For the propagators, we obtain:
\begin{eqnarray}
\begin{array}{ll}
\ds D_\phi(k) = \frac{1}{c_d} \mathcal{P}(k) \,, &
\ds D_\Phi(k) = \frac{1}{4} \mathcal{P}(k)\,,\\
\ds {[}D_A(k){]}_{ij} = - \delta_{ij} \mathcal{P}(k)  \,, &
\ds {[}D_\sigma(k){]}_{ijkl} = \tilde{\delta}_{ijkl} \mathcal{P}(k)\,,
\end{array}
\end{eqnarray}
where $\tilde{\delta}_{ijkl} \equiv \delta_{ik} \delta_{jl} + \delta_{il} \delta_{jk} + (2-c_d)\delta_{ij}\delta_{kl}$ and $\mathcal{P}(k) \equiv -(i/2)(\K^2-k_0^2)^{-1}$. The homogeneous scaling of $\mathcal{P}(k)$ with the velocity follows from the propagator expansion
\begin{equation}
\frac{1}{\K^2-k_0^2} = \frac{1}{\K^2} \sum_{n=0}^{\infty} \left( \frac{k_0^2}{\K^2}\right)^n\,,
\end{equation}
which holds in the potential region, since there we have the scaling  $k_0^2/\K^2 \sim v^2$.

Finally, Feynman diagrams can be used to organize the perturbative series, as they present a definite power counting in the PN parameter $v$. Hence, since the kinetic part of the effective action can be factored out in Eq.~\eqref{pathintegralforSTtheories} (being simply the point-particle action \eqref{eq:sourceaction} at zeroth order in the fields), Feynman diagrams will contribute to the effective two-body potential. Like in the diagrams in Fig.~\ref{fig:topologyG1G2}, two horizontal solid lines represent the two particles' worldlines, while internal lines represent gravitational and scalar interactions. In particular, only fully connected diagrams without internal loops account for classical contributions \cite{Sturani:2021ucg}.

\subsection{Spinning effects}

Now, let us include the spinning effects of the BHs. In Appendix \ref{sec:spinGRreview}, we review how spins can be consistently introduced in NRGR using the Routhian formalism. Readers unfamiliar with it are referred to this appendix before continuing. In this subsection, we shall extend this construction to the case of ESGB gravity.

The most natural extension of the dynamics of spinning objects in GR to account for additional scalar field degrees of freedom entails promoting both the four momenta $p^\mu_{\rm a}$ and spin tensors $S^{\mu\nu}_{\rm a}$ in the action \eqref{eq:ppactionspin} to scalar field-dependent quantities:
\begin{equation}
S =  \sum_{\rm a} \int d\lambda_{\rm a} \left( p^\mu_{\rm a}(\vphi) u^{\rm a}_\mu + \frac12 S_{\rm a}^{\mu\nu}(\vphi) \Omega^{\rm a}_{\mu\nu} \right)\,,
\end{equation}
where $\Omega^{\rm a}_{\mu\nu}$ is the generalized angular velocity of particle a. 
From this action, it follows that the derivation of the MPD equations at the level of the point-particle approximation remains unchanged\footnote{Indeed, it is well-known that the introduction of a scalar field modifies the MPD equations with the inclusion of finite-size terms on their right-hand sides \cite{Obukhov:2014mka}. The incorporation of such effects in our EFT framework will be discussed separately below. }. 
Hence, 
the effect of the scalar field is implicitly given through the Riemann tensor of the background geometry, which is modified from that of GR due to the presence of the scalar field.

In particular, being an algebraic relation, the use of the covariant SSC to derive an expression for $p^\mu$ in terms of $u^\mu$ and curvature corrections yields the same equation as Eq.~\eqref{eq:momentumPUR}, with the only difference being that now the mass $m = - p\cdot u$ is a function of the scalar field, {\it i.e.}, $m = m(\vphi)$. Thus, the basic Routhian for spinning particles in ESGB gravity reads: 
\begin{align}
\mathcal{R} = &\sum_{\rm a} \big( - m_{\rm a}(\vphi) \sqrt{-u_{\rm a}^2} + \frac12 S^{\rm a}_{ab}(\vphi) \omega^{ab}_{\mu} u^\mu_{\rm a}   \nonumber\\
&+ \frac{1}{2m_{\rm a}(\vphi)} R_{abcd} S^{ab}_{\rm a}(\vphi) S^{ce}_{\rm a}(\vphi) \frac{u^d_{\rm a} u^{\rm a}_e}{\sqrt{-u^2_{\rm a}}} \big) + \dots\, ,
\label{eq:routhianESGB}
\end{align}
where $\omega^{ab}_{\mu}$ are the Ricci rotation coefficients.
Two important points are worth recalling: in this formalism, spins are always given in terms of their projection in the local frame, {\it i.e.}, the indices $a,b,\dots$ are vierbein indices, and this Routhian does not include finite-size corrections, which must be added subsequently. We will address the latter point shortly.

The description above of particles moving in a scalar field environment using the point-particle action with $m \rightarrow m(\vphi)$ coincides with the prescription introduced by Eardley in 1975 for the skeletonization of compact objets in ST theories \cite{1975ApJEardley}. Since then, this prescription has been used to compute post-Newtonian corrections to the dynamics of compact binary systems in various beyond-GR theories that involve a scalar field; See for instance Refs.~\cite{Kuntz:2019zef,Diedrichs:2023foj,Bhattacharyya:2023kbh,Bernard:2023eul,Almeida:2024uph}. Nevertheless, as we will see in the following subsection, although we have introduced a scalar field dependency in the spin variables, the matching is actually realized for scalar-independent values of the spin.

For the finite-size corrections due to the rotation-induced multipole moments, the scalar field environment similarly modifies these moments, making them dependent on the scalar field:
\begin{subequations}\label{eq:finitesizeShESGB}
\begin{equation}
S^h_{\rm fs} =  -\frac12 \int d\lambda\, Q^{ab}_E(\vphi) \frac{E_{ab}}{\sqrt{-u^2}}\,, 
\end{equation}
with
\begin{equation}
Q^{ab}_E = \frac{C_{ES^2}(\vphi)}{m(\vphi)} S^a{}_c S^{bc}\,,
\end{equation}
\end{subequations}
and likewise for the other multipole moments of this series. Additionally, we must consider the effects of rotation on the scalar field multipole moments. As in pure GR, the only relevant higher-dimensional operator contributing to the 3PN order is given by the scalar-quadrupole moment coupling:
\begin{subequations}\label{eq:finitesizeSphiESGB}
\begin{equation}
S^{\vphi}_{\rm fs} = - \frac12 \int d\lambda \sqrt{-u^2}\, Q^{ab}_{\vphi}(\vphi) \nabla^\perp_a\nabla^\perp_b\vphi\,, 
\end{equation}
with
\begin{equation}
Q^{ab}_\vphi = \frac{C_{\vphi S^2}(\vphi)}{m(\vphi)} S^{ac} S^b{}_c\,,
\end{equation}
\end{subequations}
where $\nabla^\perp_a \equiv (\delta^b_a - \frac{u^b u_a}{u^2}) \nabla_b$. The orthogonatily property of this operator, {\it i.e.}, $u^a \nabla^\perp_a=0$, is crucial here since it guarantees that the SSC will still be preserved after the inclusion of the above finite-size coupling, as it happens in GR \cite{rothstein2008spin1spin1} with the orthogonality of $E_{ab}$.

Moreover, the absence of dipole, as well as octupole, moments, which would otherwise contribute to the 3PN effective potential, is justified by the symmetries of the solutions obtained in Sec.~\ref{subsec:rotBHsols}. Indeed, parity forbids couplings to these multipole moments as it would be immediately realized in the matching, which we now discuss.

\subsection{The matching}\label{sec:thematching}

In moving from the description of an isolated body at length scales $r_s \sim Gm$ to the point-particle viewpoint of a compact body in a binary system at scales $r$, a matching procedure is required to determine the Wilson coefficients of the particle's effective description. This process is essential for accurately modeling the dynamics of binaries in any given theory. 

In the case of BHs in ESGB gravity, the matching entails using information of the solutions obtained in Secs.~\ref{subsec:statbhsols} and \ref{subsec:rotBHsols} to compute the Wilson coefficients of the effective actions \eqref{eq:routhianESGB}, \eqref{eq:finitesizeShESGB}, and \eqref{eq:finitesizeSphiESGB}. These coefficients include the mass function $m(\vphi)$, spin $S^{ab}(\vphi)$, and finite-size quantities $C_{ES^2}(\vphi)$ and $C_{\vphi S^2}(\vphi)$, which will be given in terms of an appropriate set of parameters that fully characterize the BHs in the binary. As will be discussed shortly, the choice of these parameters is closely related to the thermodynamic properties of these objects.

To perform the matching, we follow \cite{BertiJulie2019} and compare the exterior fields in the large distance limit generated by an ESGB BH from two perspectives: directly solving the field equations and using the effective field description of the body as a point particle. For the latter, we begin by considering the first two terms of the Routhian in \eqref{eq:routhianESGB} for a single object:
\begin{equation}\label{eq:routhiansinglepp}
\int d\lambda_{\rm a}\,\mathcal{R}_{\rm a} = \int d\lambda_{\rm a}\,\left( - m_{\rm a}(\vphi) \sqrt{-u_{\rm a}^2} + \frac12 S^{\rm a}_{ab}(\vphi) \omega^{ab}_{\mu} u^\mu_{\rm a} \right) \,.
\end{equation}
Then, the exterior gravitational and scalar fields generated by this Routhian are obtained through the one-point functions
\begin{subequations}
	\label{eq:fieldeqs1ptfunc}
\begin{align}
\braket{h_{\mu\nu}(x)} &= \int \mathcal{D}h  \mathcal{D}\vphi\, h_{\mu\nu}(x) e^{i S[h,\vphi]}\,,\\
\braket{\vphi(x)} &= \int \mathcal{D}h \mathcal{D}\vphi\, \vphi(x) e^{i S[h,\vphi]}\,.
\end{align}
\end{subequations}
In this expression, the angle brackets represent classical corrections to the background field values $\eta_{\mu\nu}$ and $\vphi_0$, and $S[h,\vphi]$ denotes the action entering the path integral \eqref{pathintegralforSTtheories}, which includes the Einstein-Hilbert action, gauge fixing term, and an action for the above Routhian, which is treated at this level as a Lagrangian and plays the role of the matter action $S_m$. As mentioned before, we use harmonic coordinates to fix the gauge. 

\begin{figure}
\centering
	\includegraphics[scale=0.26]{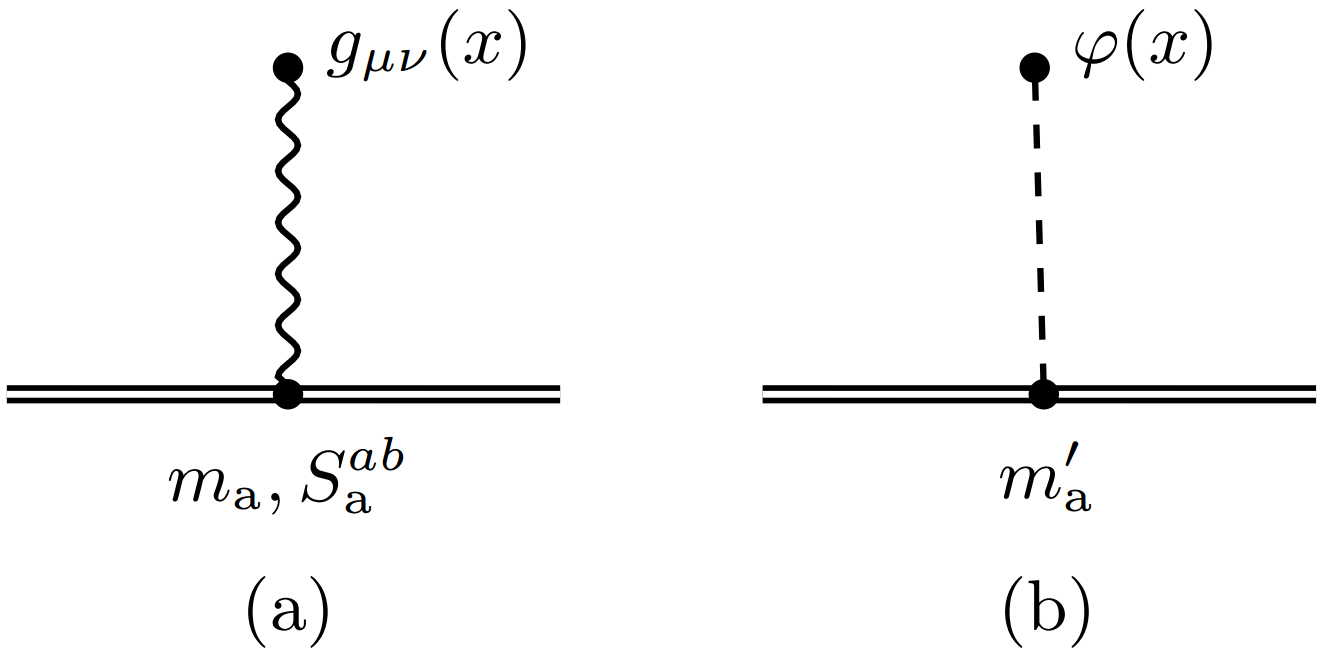}
\caption{Diagrams (a) and (b) represent leading contributions to the perturbative fields on the gravitational and scalar background values, $\eta_{\mu\nu}$ and $\vphi_0$. In (a), the wavy line represents gravitational modes sourced by either $m_{\rm a}$ or $S^{ab}_{\rm a}$, while in (b), a dashed line represents a scalar interaction sourced by $m'_{\rm a}$. The double horizontal lines represent the particle's worldline.}
\label{fig:emissiongravscalar}
\end{figure}

The leading contributions to the exterior fields generated by a pointlike object at $\X = 0$ are obtained from the computation of the diagrams in Fig.~\ref{fig:emissiongravscalar}, resulting in 
\begin{subequations}
	\label{eq:externalEFTgmunuvphi}
\begin{align}
g_{00} &= -1 + \frac{2G m_{\rm a}(\vphi_0)}{r} + \mathcal{O}\left(\frac{1}{r^2} \right)\,, \label{eq:externalEFTg00}\\
g_{0i} &= \frac{2G S^{\rm a}_{ia}(\vphi_0) n^a}{r^2} + \mathcal{O}\left(\frac{1}{r^3} \right)\,, \label{eq:externalEFTg0i} \\
g_{ij} &=\delta_{ij} \left( 1 + \frac{2G m_{\rm a}(\vphi_0)}{r} \right) + \mathcal{O}\left(\frac{1}{r^2} \right)\,, \label{eq:externalEFTgij}
\end{align}
and
\begin{equation}
\vphi = \vphi_0 - \frac{1}{r} \frac{dm_{\rm a}}{d\vphi}(\vphi_0) + \mathcal{O}\left(\frac{1}{r^2} \right)\,. \label{eq:externalEFTvphi}
\end{equation}
\end{subequations}

In the large $r$ limit, the BH solutions obtained in Sec.~\ref{subsec:rotBHsols} can be expressed in harmonic coordinates through the coordinate transformation\footnote{The Schwarzschild metric can be changed from Boyer-Lindquist to harmonic coordinates by simply making $r \rightarrow r+m$ \cite{kerrharmoniccoords}. When rotation is included ({\it i.e.}, the Kerr spacetime), this transformation acquires $\mathcal{O}(1/r)$ corrections. Scalar interations, on the other hand, modify this with additional $\mathcal{O}(1/r^2)$ terms.}
$r \rightarrow r + m + \mathcal{O}(1/r)$\,. For these solutions, the angular momentum is aligned with the $z$-direction, {\it i.e.}, $\theta = 0$.
Then, changing the orientation of the angular momentum to an arbitrary direction (denoted by vector ${\bf J}$ with magnitude $|{\bf J}| = J$), and introducing the dual three-tensor $J^{ij} \equiv \epsilon^{ijk}J_k$, these solutions become:
\begin{subequations}
	\label{eq:externalEEgmunuvphi}
\begin{align}
g_{00} &= -1 + \frac{2G m}{r} + \mathcal{O}\left(\frac{1}{r^2} \right)\,, \label{eq:externalEEg00}\\
g_{0i} &= \frac{2G J_{ia} n^a}{r^2} + \mathcal{O}\left(\frac{1}{r^3} \right)\,, \label{eq:externalEEg0i}\\
g_{ij} &=\delta_{ij} \left( 1 + \frac{2G m}{r} \right) + \mathcal{O}\left(\frac{1}{r^2} \right)\,, \label{eq:externalEEgij}
\end{align}
and
\begin{equation}
\tilde{\vphi}  = \vphi_0 + \frac{D}{r} + \mathcal{O}\left(\frac{1}{r^2} \right)\,. \label{eq:externalEEvphi}
\end{equation}
\end{subequations}

Comparing Eqs.~\eqref{eq:externalEFTgmunuvphi} and \eqref{eq:externalEEgmunuvphi}, we obtain the following matching conditions:
\begin{subequations}
	\label{eq:matchingconds}
\begin{align}
m_{\rm a}(\vphi_0) &= m\,, \\
m'_{\rm a}(\vphi_0) &= -D\,,\\
S_{\rm a}(\vphi_0) &= J\,, \label{eq:mathingconditionIII}
\end{align}
\end{subequations}
where, likewise, $S_a(\vphi)$ is the magnitude of the vector ${\bf S}_{\rm a}$, defined by $S^{\rm a}_i = \frac12 \epsilon_{ijk}S_{\rm a}^{jk}$. 

As discussed in Ref.~\cite{BertiJulie2019} in the context of non-rotating BHs in ESGB gravity, the first two conditions in Eq.~\eqref{eq:matchingconds} ensure that the point-particle action arising from the matching 
produces fields which equal to the BH's at all orders in a $1/r$ expansion. 
This is because the asymptotic behavior of \eqref{eq:matchingconds} can be seen as boundary conditions that identify the external fields to a unique solution of the vacuum equations of motion \eqref{eq:fieldeqsESGB} (See also Ref.~\cite{Julie:2022huo}). When rotation is included, this reasoning still holds, provided that, in addition to impose the third condition \eqref{eq:mathingconditionIII}, couplings to the spin-induced multipole moments are added to the effective point-particle action, which must also be matched te the BH solutions. This will be discussed below, after studying some important consequences of the matching conditions \eqref{eq:matchingconds}.

As we have seen in Sec.~\ref{Sec:ESGBgravity}, BHs in ESGB gravity exhibit secondary-type hair, meaning that the scalar charge is not an independent parameter but is related to the mass $m$ and scalar background value $\vphi_0$, and, in the case of rotation, also a function of the angular momentum $J$. The angular momentum, on the other hand, is unconstrained. This indicates that, at least in the Einstein frame, the spin variable appearing at the point-particle Rothian \eqref{eq:routhiansinglepp} does not develop a scalar field dependence during the skeletonization\,\footnote{This discussion is fully consistent with what is observed in ST theories for slowly-rotating neutron stars \cite{Kuntz:2024jxo}.}.
 However, this property does not hold in the Jordan frame, where the vierbein acquires a field-dependent coupling via $\tilde{e}^a_\mu = \mathcal{A}(\vphi) e^a_\mu$ as the Jordan frame metric is $\tilde{g}_{\mu\nu} = \mathcal{A}^2(\vphi) g_{\mu\nu}$. This leads to $\tilde{S}_{ab}(\vphi) \equiv \mathcal{A}(\vphi)^{-2} S_{ab}$ and induces the following
\begin{equation}
\omega_\mu^{ab} \rightarrow \tilde{\omega}_\mu^{ab} + \left( \tilde{e}^a_{\mu} \tilde{e}^{b\nu} - \tilde{e}^b_{\mu} \tilde{e}^{a\nu} \right) \mathcal{A}^{-1} \partial_\nu \mathcal{A} \,,
\end{equation}
where $\tilde{\omega}_\mu^{ab} \equiv \tilde{e}^{b\nu} \tilde{\nabla}_\mu \tilde{e}^a_\nu$, with $\tilde{\nabla}_\mu$ being the covariant derivative compatible with $\tilde{g}_{\mu\nu}$. The second term in this coupling introduces contributions to $\vphi$ that are linear in the spin, which is absent in our formulation of ESGB gravity (See Eq.~\eqref{eq:scalarfieldwspin}).

\subsubsection{Matching for mass function}

From the matching conditions \eqref{eq:matchingconds}, we see that the field-independent spin is related to the BH's angular momentum by $S_{\rm a} = J$, while an expression for $m(\vphi)$ is obtained by integrating the second equation with $m = m(\vphi_0)$ inserted into $D$ in Eq.~\eqref{eq:scalarchargeDsol}, leading to $m'(\vphi_0) = - D(m(\vphi_0),\vphi_0)$. This results in a first-order differential equation for the mass function:
\begin{align}
&\frac{m'(\varphi )}{m(\vphi)}
+\epsilon_{\rm a}(\vphi) \left(2 -\frac{\chi_{\rm a}^2(\vphi)}{2} -\frac{\chi_{\rm a}^4(\vphi)}{4} \right)   +\epsilon ^2_{\rm a}(\vphi) 
\nonumber\\
 &\cdot\left(
\frac{73 f''(\vphi )}{30 f'(\vphi)}
-\frac{21 \chi_{\rm a}^2(\vphi) f''(\vphi )}{20 f'(\vphi )}
-\frac{20687 \chi^4_{\rm a}(\vphi) f''(\vphi )}{100800 f'(\vphi )} 
\right) &
\nonumber\\
&~~ + \dots = 0\,,
\end{align}
where $\epsilon_{\rm a}(\vphi) \equiv \alpha f'(\vphi)/(4m_{\rm a}(\vphi)^2)$ and $\chi_{\rm a}(\vphi) \equiv S_{\rm a}/m({\vphi})^2$.
Integration of this equation can be performed perturbatively, and results in\footnote{In a binary system, the background value $\vphi_0$, interpreted as the asymptotic field generated by the BH companion, becomes dynamical. For this reason, the solution for the Wilson coefficients can be given in terms of the field $\vphi$, which is needed for modeling the PN dynamics of binary systems in ESGB gravity.} \\
\begin{equation}
m_{\rm a} (\vphi) = \mu_{\rm a}   \Bigg[ 1+ \sum_{n=1}^7 \sum_{l =0,2,4} \frac{\alpha^n S_{\rm a}^l F^{(n)}_l(\vphi)}{\mu^{2n+2 l}_{\rm a}} \Bigg]\,,
\end{equation}
which depends on a new integration constant $\mu_{\rm a}$ whose interpretation will be discussed below. The coefficients $F^{(n)}_l$ are functions of $\vphi$, given by the following expression (shown here up to $\mathcal{O}(\alpha^4))$:
\begin{align}
F_0^{(1)}(\vphi) &= - \frac{f(\vphi)}{2} \,, \\
F_0^{(2)}(\vphi) &= - \frac{f(\vphi)^2}{8} - \frac{73 f'(\vphi)^2}{960} \,, \\
F_0^{(3)}(\vphi) &= - \frac{f(\vphi)^3}{16} - \frac{73 f(\vphi) f'(\vphi)^2}{640} 
\nonumber  \\
&~~~~ - \frac{12511 f'(\vphi)^2 f''(\vphi)}{483840}\,, \\
F_0^{(4)}(\vphi) &= - \frac{5 f(\vphi)^4}{128} - \frac{73 f(\vphi)^2 f'(\vphi)^2}{512} 
\nonumber  \\
&~~~~ - \frac{12534857 f'(\vphi)^4}{425779200} - \frac{12511 f(\vphi) f'(\vphi)^2 f''(\vphi)}{193536} 
\nonumber  \\
& ~~~~ - \frac{227192473  f'(\vphi)^2 f''(\vphi)^2}{25546752000}
\nonumber  \\
& ~~~~ - \frac{799607 f'(\vphi)^3 f^{(3)}(\vphi)}{255467520} \,,
\end{align}
for $l = 0$,
\begin{align}
F_2^{(1)}(\vphi) &= \frac{f(\vphi)}{8} \,, \\
F_2^{(2)}(\vphi) &= \frac{3f(\vphi)^2}{16} + \frac{21 f'(\vphi)^2}{640} \,, \\
F_2^{(3)}(\vphi) &= \frac{15 f(\vphi)^3}{64} + \frac{367 f(\vphi) f'(\vphi)^2}{2560}
\nonumber\\
&~~~~ + \frac{ 3206149 f'(\vphi)^2 f''(\vphi)}{212889600} \,,\\
F_2^{(4)}(\vphi) &= \frac{35 f(\vphi)^4}{128} + \frac{917 f(\vphi)^2 f'(\vphi)^2}{2560} + \frac{82248191 f'(\vphi)^4}{82248191} \nonumber  \\
&~~~~ +  \frac{ 35736391 f(\vphi) f'(\vphi)^2 f''(\vphi)}{425779200} 
\nonumber  \\
& ~~~~ +\frac{4301149469 f'(\vphi)^2 f''(\vphi)^2}{664215552000}
\nonumber  \\
&~~~~ + \frac{148901437 f'(\vphi)^3 f^{(3)}(\vphi)}{66421555200}\,,
\end{align}
for $l = 2$, and 
\begin{align}
F_4^{(1)}(\vphi) &= \frac{f(\vphi)}{16} \,, \\
F_4^{(2)}(\vphi) &= \frac{15f(\vphi)^2}{128} + \frac{20687 f'(\vphi)^2}{3225600} \,, \\
F_4^{(3)}(\vphi) &= \frac{43 f(\vphi)^3}{256} + \frac{134317 f(\vphi) f'(\vphi)^2}{6451200}
\nonumber  \\
&~~~~ - \frac{ 60900137 f'(\vphi)^2 f''(\vphi)}{34594560000} \,,\\
F_4^{(4)}(\vphi) &= \frac{217 f(\vphi)^4}{1024} + \frac{ 125131 f(\vphi)^2 f'(\vphi)^2}{25804800} 
\nonumber  \\
&~~~~ - \frac{26709048987713 f'(\vphi)^4}{2712213504000000} 
\nonumber  \\
&~~~~ -  \frac{ 864532121 f(\vphi) f'(\vphi)^2 f''(\vphi)}{42577920000} \nonumber  \\
&~~~~ -\frac{196393841309 f'(\vphi)^2 f''(\vphi)^2}{82188288000000}
\nonumber  \\
&~~~~ - \frac{ 91508684921 f'(\vphi)^3 f^{(3)}(\vphi)}{92990177280000}\,,
\end{align}
for $l = 4$. 
The remaining expressions for up to $\mathcal{O}(\alpha^7)$ can be found in the ancillary Mathematica notebook of the supplemental material.

The coefficients for $l = 0$ have been previously obtained for the non-spinnning case up to $n=4$ \cite{BertiJulie2019}, and we find complete agreement for these functions. On the other hand, the coefficients for $l = 2$ and $l=4$ are only present when rotation is considered, and, to the best of our knowledge, is derived here for the first time.

The physical interpretation of the parameter $\mu_{\rm a}$ was thoroughly explored in Ref.~\cite{BertiJulie2019} in the context of non-rotating BHs in ESGB gravity, which was show to correspond to the irreducible mass, $\mu_{\rm a} = M_{\rm irr}$, related to the Wald entropy by $M_{\rm irr} = \sqrt{S_{\rm w}/4\pi}$. When rotation is included, the relation is now given by:
\begin{equation}\label{eq:relationSwmuJ}
\frac{S_{\rm w}}{4\pi} =  \mu_{\rm a}^2 - \frac{S_{\rm a}^2}{4\mu_{\rm a}^2} - \frac{S_{\rm a}^4}{16\mu_{\rm a}^6} + \mathcal{O}(S_{\rm a}^6)  \,.
\end{equation} 
This is simply an spin expansion of the well-known general relativistic irreducible mass \cite{Wald:1984rg}
\begin{equation}
M_{\rm irr}^2  = \frac12 \left[ \mu_{\rm a}^2  + ( \mu_{\rm a}^4 - S_{\rm a}^2)^{\frac12} \right]\, .
\end{equation} 
That is, in this case, $\mu_{\rm a}$ actually represents the relativistic mass of the BH, which is only related to the irreducible mass through the relation above. In any case, the discussions from Ref.~\cite{BertiJulie2019} can be easily extended to the spinnning case, since here the matching conditions also force the BH entropy to be constant, $\delta S_{\rm w}=0$. Thus, the mass parameter $\mu_{\rm a}$, which is implicitly related to the thermodynamic quantities $S_{\rm w}$ and $S_{\rm a}$ through Eq.~\eqref{eq:relationSwmuJ}, and the spin magnitude $S_{\rm a}$ are suitable parameters to characterize the BH, as they remain unchanged during the bodies' slow inspiral. In contrast, the ADM mass and scalar charge of the BH, being functions of $\vphi_0$, evolve adiabatically through equilibrium configurations ({\it i.e.}, maintaining constant Wald entropy) during the system's evolution. 

\subsubsection{Matching for finite-size couplings}

We now turn to the matching of the coefficients $C_{ES^2}(\vphi)$ and $C_{\vphi S^2}(\vphi)$. This can be accomplished by computing the one-point functions $\braket{h_{00}(x)}$ and $\braket{\vphi(x)}$ using Eq.~\eqref{eq:fieldeqs1ptfunc}, with contributions generated by the finite-size operators $S^h_{\rm fs}$ and $S^\vphi_{\rm fs}$ in Eqs.~\eqref{eq:finitesizeShESGB} and \eqref{eq:finitesizeSphiESGB}. This entails evaluating the diagrams shown in Fig.~\ref{fig:emissiongravscalarfs}, resulting in the following contributions:
\begin{figure}
\centering
	\includegraphics[scale=0.26]{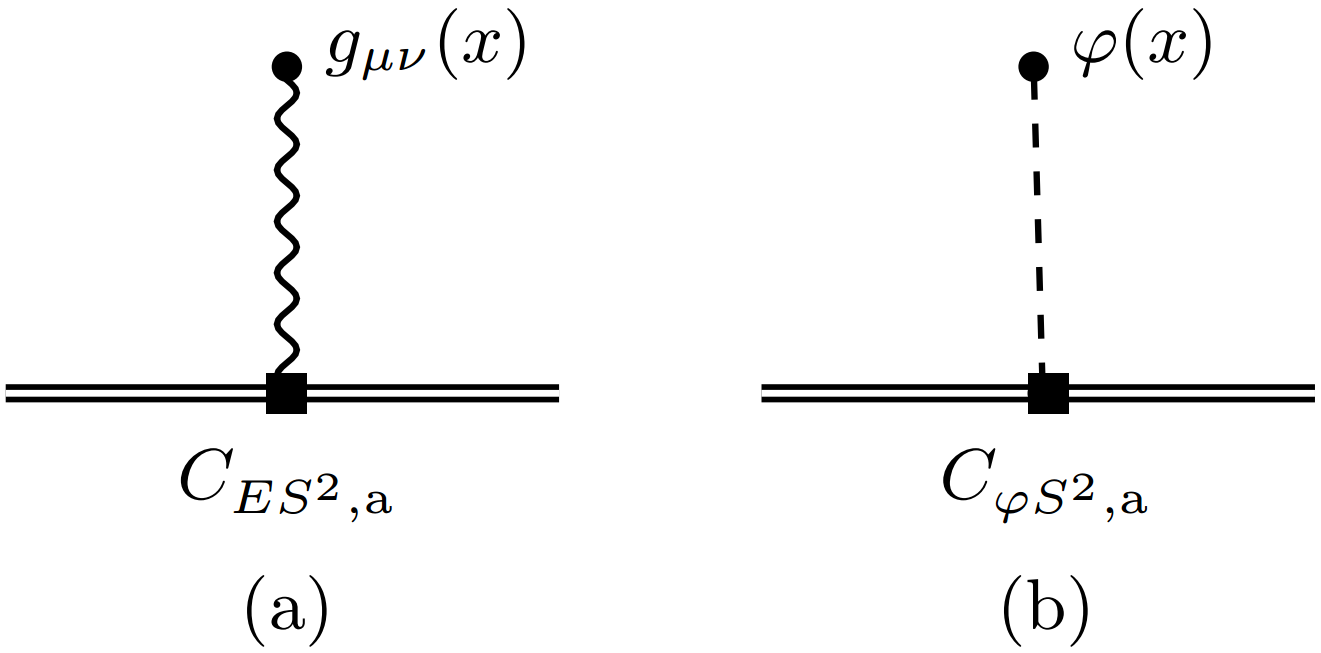}
\caption{Diagrams (a) and (b) depict contributions to the exterior gravitational and scalar fields generated by the finite-size couplings $C_{ES^2,{\rm a}}$ and $C_{\vphi S^2,{\rm a}}$, each of which is represented by a black square.}
\label{fig:emissiongravscalarfs}
\end{figure}
\begin{align}
\braket{h_{00}(x)} &= \frac{G C_{ES^2,{\rm a}}(\vphi_0)}{m_{\rm a}(\vphi_0) r^3} \left( \Ss^2_{\rm a} - 3(\Ss_{\rm a}\cdot \RR)^2 \right) \,, \label{eq:g00quadrmoment0} \\
\braket{\vphi(x)} &=  - \frac{G C_{\vphi S^2,{\rm a}}(\vphi_0)}{2m_{\rm a}(\vphi_0) r^3} \left( \Ss^2_{\rm a} - 3(\Ss_{\rm a}\cdot \RR)^2 \right) \,. \label{eq:vphiquadrmoment0}
\end{align}
These results can then be directly compared with the quadrupole parts of Eqs.~\eqref{eq:gttquad} and \eqref{eq:vphiquad}, taking into account the matching conditions \eqref{eq:matchingconds}, whence it follows
\begin{align}
C_{ES^2,{\rm a}}(\vphi) &= 1 + \frac{17852}{2625}\epsilon^2_{\rm a}(\vphi) + \dots \,, \label{eq:gravcouplingmatching}\\
C_{\vphi S^2,{\rm a}}(\vphi) &= - \frac{38}{15}\epsilon_{\rm a}(\vphi) - \frac{15853 f''(\vphi)}{5250 f'(\vphi)} \epsilon^2_{\rm a}(\vphi) + \dots\,.
\end{align}
The complete expressions valid for up to $\mathcal{O}(\epsilon_{\rm a}^7)$ is provided in the ancillary Mathematica notebook.

It is important to note that the direct comparison of expressions \eqref{eq:gttquad} and \eqref{eq:vphiquad} with \eqref{eq:g00quadrmoment0} and \eqref{eq:vphiquadrmoment0} is only possible because the coordinate system used in Sec.~\ref{subsubssec:geomproperties} to extract the quadrupole moments fully coindices with the harmonic coordinates in the asymptotic expansion of the fields \cite{Thorne:1980ru}.

\section{Post-Newtonian dynamics of spinning binaries}\label{sec:3PNdynamicsESGB}

Now that the effective point-particle description of a BH in ESGB gravity has been extended to the spinning case, in a procedure which is equivalent to the ``skeletonization" \cite{BertiJulie2019,1975ApJEardley}, we are in position to study the PN dynamics of BH binaries in this theory. Thus, in this section, we investigate the additional contributions needed to supplement existing results from GR, ST theories, and ESGB gravity, to provide a complete description of the dynamics of spinning binaries in ESGB gravity up to the 3PN level. Notably, the 3PN effective potential for the non-spinning case is fully accounted for by the mass term in the effective point-particle Routhian \eqref{eq:routhianESGB}, as previously obtained in \cite{Julie:2022qux}. Therefore, we only need to consider new diagrams involving spin couplings that include at least one scalar propagator. Additionally, for now, we will ignore contributions from the Gauss-Bonnet coupling in the action $S_{\vphi}$ in Eq.~\eqref{phiaction0}, postponing the discussion for Sec.~\ref{sec:gbcouplingout}.

Here we use the EFT approach discussed in Sec.~\ref{sec:spinsinESGBgrav}, valid for compact binaries on bound orbits, with point-particle Routhian given by Eq.~\eqref{eq:routhianESGB} and finite-size couplings by Eqs.~\eqref{eq:finitesizeShESGB} and \eqref{eq:finitesizeSphiESGB}, and work in harmonic coordinates. Also, we reintroduce the Newton's constant $G$, as this is convenient to power count the PN order of the contributing diagrams. 
Thus, below we compute all the necessary contributions to the effective two-body potential to complete the 3PN order dynamics of spinning BH binaries in ESGB gravity.
\begin{figure}
\centering
	\includegraphics[scale=0.26]{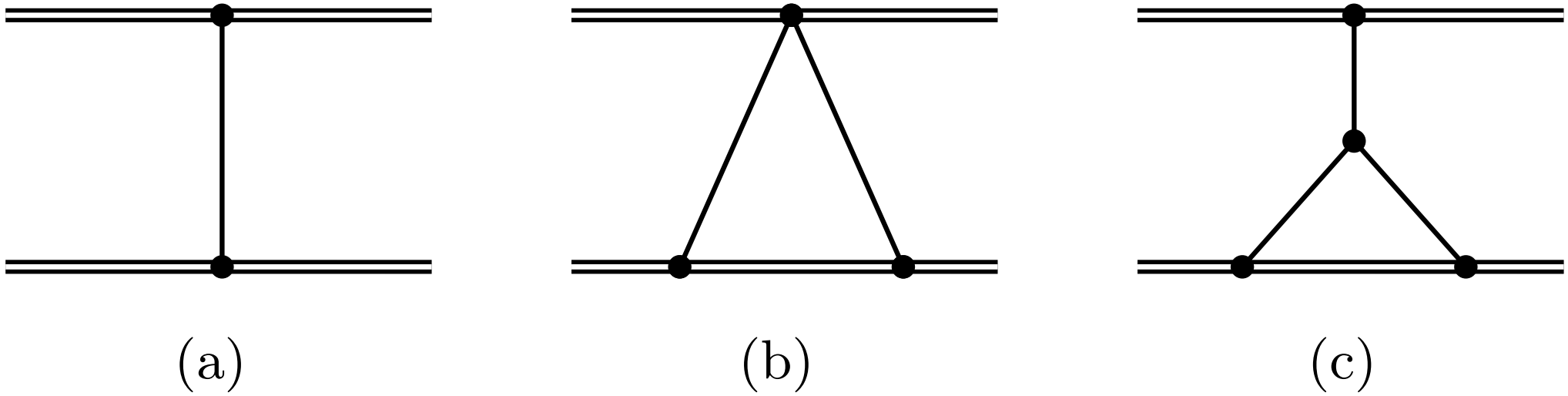}
\caption{Diagram topologies that contribute with spin corrections to the 3PN conservative dynamics of a binary system. The two horizontal double lines represent the worldlines of the two objects, while the internal lines, depicted generically by solid lines, represent potential exchanges (propagators) of either gravitational or scalar nature. Diagram (a) is the only $\mathcal{O}(G)$ topology. In contrast, diagrams (b) and (c), which correspond to the ``seagull" topology and the one with internal vertex, respectively, are of $\mathcal{O}(G^2)$.}
\label{fig:topologyG1G2}
\end{figure}

\subsection{Minimal coupling}

We begin by considering the minimal sector of the point-particle Routhian \eqref{eq:routhianESGB}, which is given by: 
\begin{equation}\label{eq:mincouplingresults}
\mathcal{R} = \sum_{\rm a} \left( - m_{\rm a}(\vphi) \sqrt{-u_{\rm a}^2} + \frac12 S^{\rm a}_{ab} \omega^{ab}_{\mu} u^\mu_{\rm a} \right)\,.
\end{equation}
\begin{figure}
\centering
	\includegraphics[scale=0.26]{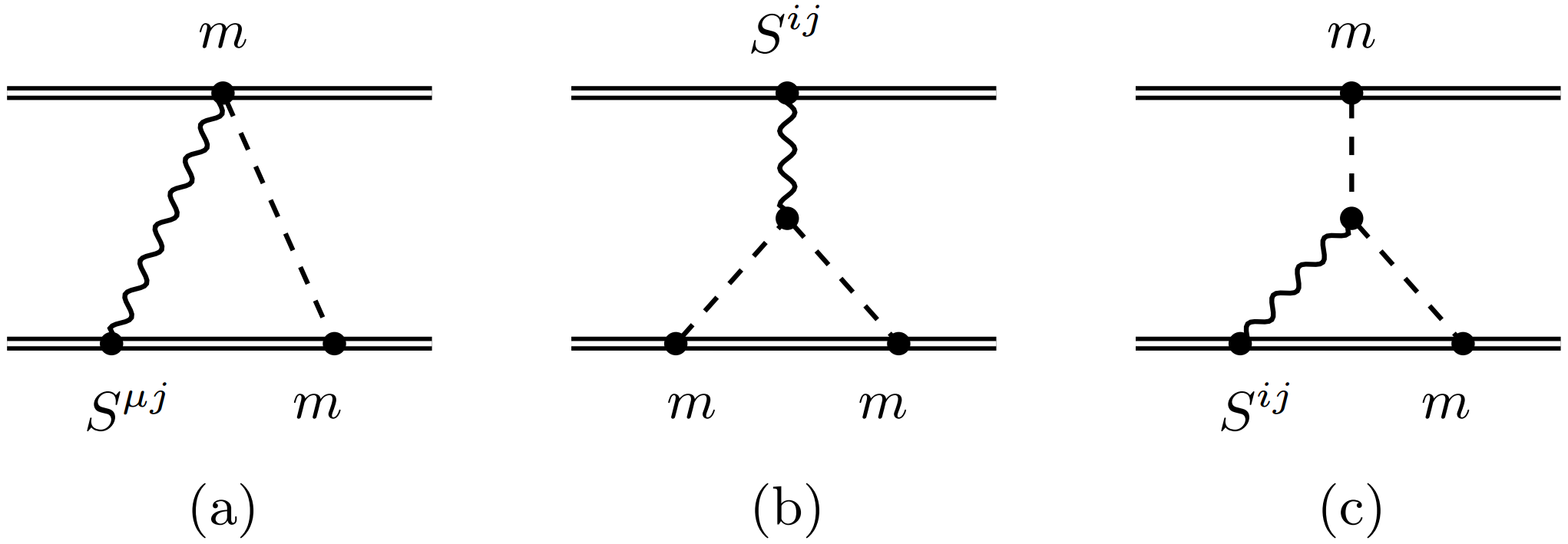}
\caption{Diagrams contributing with spin-orbit effects to the effective two-body potential, arising from the minimal coupling \eqref{eq:mincouplingresults}. Diagram (a) actually represents three distinct diagrams: one for $\mu = 0$ involving a gravitational mode $\phi$, and two with $\mu = i$, one with a gravitational mode $\phi$ and the other with $A_i$. Similarly, each of diagram (b) and  diagram (c) includes two diagrams, having an exchange of either $A_i$ or  $\sigma_{ij}$ gravitational mode. }
\label{fig:SSG2SS}
\end{figure}

Note that no diagrams of the $\mathcal{O}(G)$ topology in Fig.~\ref{fig:topologyG1G2}(a) can include both scalar modes and spin couplings, as $S_{ab}$ does not depend on $\vphi$. Therefore, such a diagram topology cannot contribute. Instead, new contributions to the 3PN level in this sector arise from the two diagrams of $\mathcal{O}(G^2)$ depicted in Figs.~\ref{fig:topologyG1G2}(b) and \ref{fig:topologyG1G2}(c).

For the ``seagull" topology in Fig.~\ref{fig:topologyG1G2}(b), only couplings linear in the spin can contribute. A quadratic coupling would involve only graviton modes and is thus fully accounted for in the pure GR case. Similarly, considering two linear spin couplings in this diagram topology would also only involve graviton modes. Therefore, additional diagrams with this topology can only contribute to spin-orbit effects. This conclusion extends to the other topology in Fig.~\ref{fig:topologyG1G2}(c), as the three-scalar-graviton vertices derived from $S_{\vphi}$ contain two scalar modes, which can only be connected to the mass coupling in Eq.~\eqref{eq:mincouplingresults}.

The diagrams that contribute in this case are shown in Fig.~\ref{fig:SSG2SS}. In particular, the first of them, Fig.~\ref{fig:SSG2SS}(a), contains three subdiagrams contributing at the 2.5PN order, whose sum results in the following correction to the two-particle potential:
\begin{equation}\label{eq:mincouplingseagullresult}
V^{\rm (a)}_{0} = \frac{G^2 m^0_1 m^0_2}{r^3}\alpha^0_1\alpha^0_2 n^i \left[ S_1^{0i} - S_1^{ij} (v_1^j - 2 v_2^j) \right] + (1 \leftrightarrow 2)\,.
\end{equation}
Likewise, for the topology involving a three-scalar-graviton vertex, we have a total of four subdiagrams contained in Figs.~\ref{fig:SSG2SS}(b) and \ref{fig:SSG2SS}(c), each, if non-vanishing, contributing at 2.5PN order:
\begin{align}
V^{\rm (b)}_{0} &= -\frac{G^2 (m^0_2)^2}{2r^3}(\alpha^0_2)^2 S_1^{ij} n^i (v_1^j - v_2^j) + (1 \leftrightarrow 2)\,,\\
V^{\rm (c)}_{0} &= 0\,.
\end{align}
Note that the result for $V^{(a)}_0$ in \eqref{eq:mincouplingseagullresult} is just the GR result multiplied by $\alpha^0_1\alpha^0_2$ \cite{Porto:2005ac}. This occurs because the worldline couplings to $\phi$ and $\vphi$ differ only by a factor of $\alpha^0_{\rm a}$ and because the relevant quadratic vertices are proportional to $\phi^2/2$ and $\phi\vphi$, with has a factor of 2 compensated by the symmetry factor of the seagull topology.

\subsection{Finite-size coupling to gravity}

Now we consider the gravitational finite-size coupling $S^h_{\rm fs}$ in \eqref{eq:finitesizeShESGB}, with the coefficient expanded according to 
\begin{equation}\label{eq:Wilsoncoefffsgrav}
	\frac{C_{ES^2,{\rm a}}(\vphi)}{m_{\rm a}(\vphi)} = \frac{1}{m^0_{\rm a}} \sum_{n = 0}^{\infty} C^{(n)}_{ES^2,{\rm a}} (\vphi-\vphi_0)^n\,.
\end{equation}
The constant coefficients $C^{(n)}_{ES^2,{\rm a}}$ can be thought of as spin-induced quadrupole sensitivities, which measure the strength of the body's quadrupole moment in the presence of an scalar field environment. 

To 3PN order, only the first two coefficients in Eq.~\eqref{eq:Wilsoncoefffsgrav} are relevant. For the $n=0$ term, the contributions originally computed in GR for spin-induced finite-size effects \cite{rothstein2008spin1spin1} can be adapted by replacing the Wilson coefficient $C_{ES^2}$ with $C_{ES^2}^{(0)}$. This results in contributions to the two-body potential beginning at the 2PN order.

For the $n=1$ case, new diagrams involving scalar modes emerge at the $\mathcal{O}(G^2)$ level, which contribute at the 3PN order. These contributions are given by the two diagrams in Fig.~\ref{fig:SOG2linSfs} and yield:
\begin{figure}
\centering
	\includegraphics[scale=0.26]{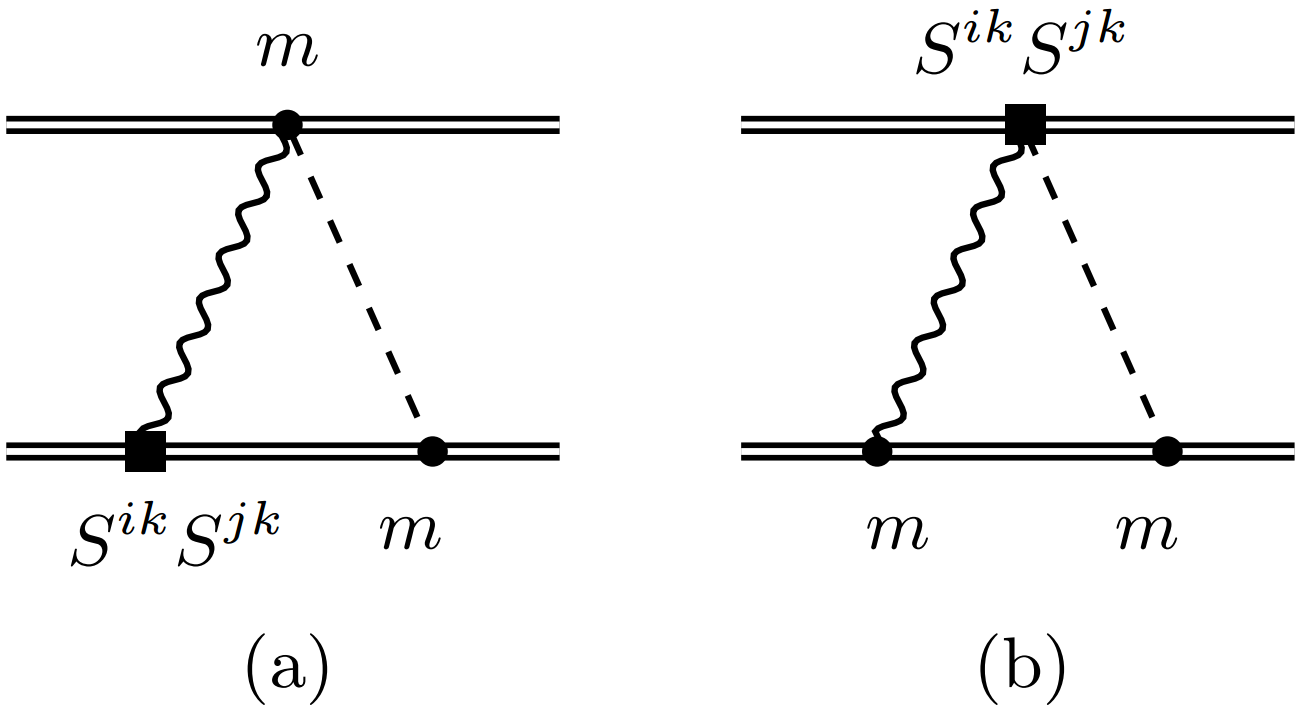}
\caption{In (a) and (b), we show the only additional diagrams that must be computed here. These diagrams account for rotation-induced finite-size corrections from the coupling with gravity, involving scalar interactions.}
\label{fig:SOG2linSfs}
\end{figure}
\begin{align}
V_{h,{\rm fs}}^{\rm (a)} &= - \frac{G^2 m^0_2 \alpha^0_1 \alpha^0_2}{2r^4} C^{(0)}_{ES^2,1} S^{ik}_1 S^{jk}_1 (\delta_{ij} - 3n_i n_j) \nonumber\\ 
&+ (1\leftrightarrow 2)\,,\\
V_{h,{\rm fs}}^{\rm (b)} &= - \frac{G^2 (m^0_2)^2 \alpha^0_2}{2 m^0_1 r^4}  C^{(1)}_{ES^2,1} S^{ik}_1 S^{jk}_1 (\delta_{ij} - 3n_i n_j) \nonumber\\ 
&+ (1\leftrightarrow 2)\,.
\end{align}
In both diagrams shown in Fig.~\ref{fig:SOG2linSfs}, the gravitational modes are given only by the $\phi$ field. Exchanges involving $A_i$ or $\sigma_{ij}$ would introduce additional powers of $v$ from the worldline couplings, leading to contributions at higher orders. Diagrams with a three vertex are disregarded here since they involve time derivatives from the $\phi\vphi^2$ term, which again pushes the scaling in $v$, only leading to contributions beyond the 3PN order.

\subsection{Finite-size coupling to scalar field}\label{subsec:fsgravcoupling}

Similarly to the previous case, we expand the coefficient of $S^{\vphi}_{\rm fs}$ in Eq.~\eqref{eq:finitesizeSphiESGB} as follows:
\begin{equation}\label{eq:Wilsoncoefffsscalar}
	\frac{C_{\vphi S^2,{\rm a}}(\vphi)}{m_{\rm a}(\vphi)} = \frac{1}{m^0_{\rm a}} \sum_{n = 0}^{\infty} C^{(n)}_{\vphi S^2,{\rm a}} (\vphi-\vphi_0)^n\,.
\end{equation}
In this case, we must consider not only the new contributions from diagrams with $\mathcal{O}(G^2)$ topologies but also those of $\mathcal{O}(G^1)$. This is because $S_{\rm fs}^\vphi$ directly couples the spin to the scalar field $\vphi$. In particular, for quadratic terms involving a graviton and a scalar mode, care must be taken as the indices $a,b$ in $Q^{ab}_{\vphi}$ are vierbein indices. Taking this into consideration, the relevant expansion of  the action $S_{\rm fs}^\vphi$ we need is given by
\begin{align}
S^\vphi_{\rm fs} &= -\frac1{2 m^0 \Lambda^2} \int dt\,  S^{ik} S^j{}_k \times \nonumber\\
&\bigg\{ \Lambda C^{(0)}_{\vphi S^2} \bigg[ \left(  1-\frac{v^2}{2}  \right)  \partial_i\partial_j\Phi
+ 2 v_j v^l \partial_i\partial_l\Phi \bigg] \nonumber \\
&+ C^{(1)}_{\vphi S^2} \Phi \partial_i\partial_j\Phi +  C^{(0)}_{\vphi S^2}  (c_d-1)  \phi   \partial_i\partial_j\Phi  \nonumber\\ 
&+ \frac12 (c_d - 2)  C^{(0)}_{\vphi S^2} (3\partial_i\phi \partial_j\Phi - \delta_{ij} \partial_l\phi \partial_l\Phi)  
\bigg\}\,.
\end{align}
Here, recall that $\Phi$ represents the potential modes introduced in Sec.~\ref{sec:EFTsetup}.

\begin{figure}
\centering
	\includegraphics[scale=0.26]{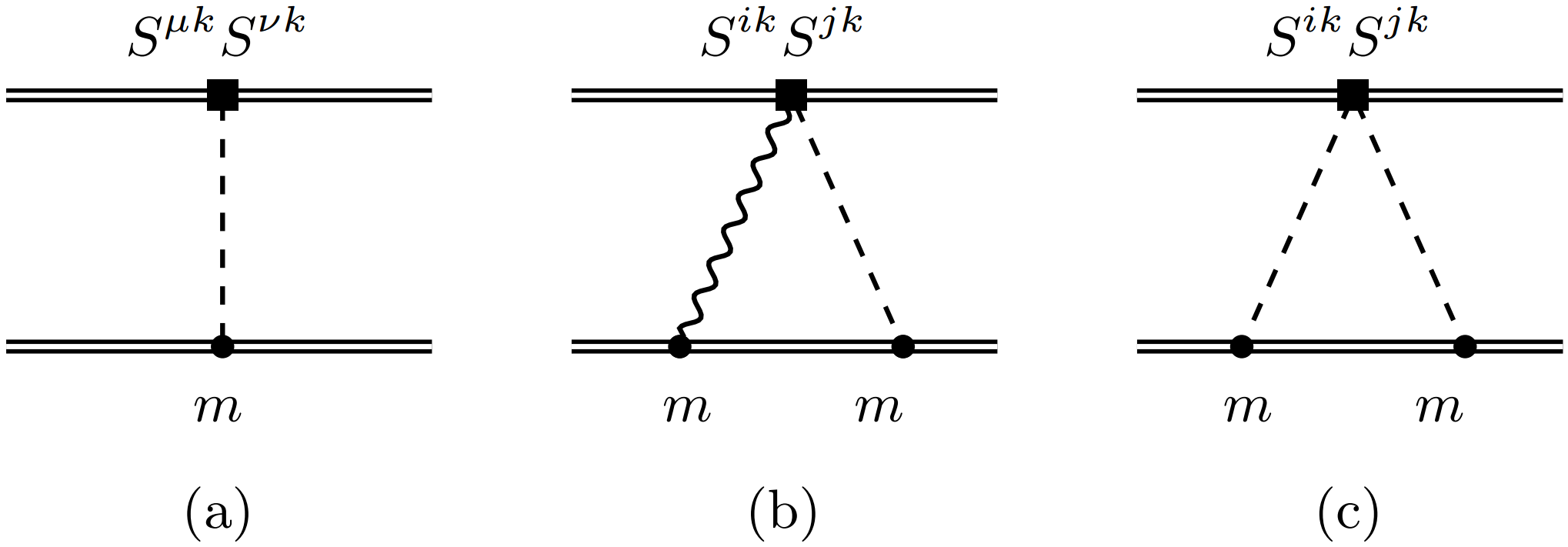}
\caption{In (a), (b), and (c), we show the additional diagrams accounting for rotation-induced finite size corrections, from the coupling to the scalar field, that must be considered. In particular, the diagram topology in (a) gives rise to five contributing diagrams, each for the different possibilities of indices $\mu$ and $\nu$, and also in the powers in $v$ of the two vertices. On the other hand, the diagrams in (b) and (c) contribute with one single diagram each. }
\label{fig:topologyG1G2fsscalar}
\end{figure}

In this case, we start by computing the diagram in Fig.~\ref{fig:topologyG1G2fsscalar}(a), which actually represents seven distinct diagrams. These diagrams differ from one another by powers of $v$ in their vertices, as well as for the possible values $\mu$ and $\nu$ might assume in $S^{\mu k} S^{\nu k}$, being either a time or a space type of index. Summing the contributions from these diagrams yields:
\begin{align}
V^{\rm (a)}_{\vphi,{\rm fs}} &= \frac{G m^0_2 \alpha^0_2}{m^0_1 r^3} C^{(0)}_{\vphi S^2,1} S^{ik}_1 S^{jk}_1  v^i_1 [ v^j_1 - 3(\V_1\cdot \RR) n^j ] \nonumber\\
&+ \frac{G m^0_2 \alpha^0_2}{2m^0_1 r^3} C^{(0)}_{\vphi S^2,1} \bigg[ \left( 1- \frac{v_1^2}{2}- \frac{v_2^2}{2} \right) S^{ik}_1 S^{jk}_1 \nonumber\\
&- ( S^{i0}_1 S^{j0}_1 + 2 S^{0k}_1 S^{ik}_1 v^j_2) \bigg] (\delta_{ij} - 3 n_i n_j) \nonumber\\
&+ (1\leftrightarrow 2)\,,
\end{align}
which provides contributions of order 2PN and 3PN to the effective two-body potential. 
For the $\mathcal{O}(G^2)$ diagrams shown in Fig.~\ref{fig:topologyG1G2fsscalar}(b) and \ref{fig:topologyG1G2fsscalar}(c), we find:
\begin{align}
V^{\rm (b)}_{\vphi,{\rm fs}} &= - \frac{2G^2 (m^0_2)^2 \alpha^0_2}{r^4} \frac{ C^{(0)}_{\vphi S^2,1} }{m^0_1} S^{ik}_1 S^{jk}_1 (\delta_{ij} - 3 n_i n_j) \nonumber\\
&+ (1\leftrightarrow 2)\,, \\
V^{\rm (c)}_{\vphi,{\rm fs}} &= - \frac{G^2 (m^0_2)^2 (\alpha^0_2)^2}{2r^4} \frac{C^{(1)}_{\vphi S^2,1}}{ m^0_1 }  S^{ik}_1 S^{jk}_1 (\delta_{ij} - 3 n_i n_j) \nonumber\\
&+ (1\leftrightarrow 2)\,, 
\end{align}
both of which contribute to the effective potential at the 3PN level.

Similar to the gravitational case with the coupling $S^h_{\rm fs}$, diagrams featuring an internal three-vertex contribute beyond the 3PN order. This is due to the presence of time derivatives in the three vertex or additional velocity terms in the worldline coupling, which arise whenever the fields $A_i$ or $\sigma_{ij}$ are involved. Consequently, such diagrams can be disregarded here.

\subsection{Internal vertex from Gauss-Bonnet coupling}\label{sec:gbcouplingout}

In this subsection, we examine spin effects arising from processes involving the Gauss-Bonnet coupling $\propto \alpha \sqrt{-g}f(\vphi)\mathcal{G}$, given in Eq.~\eqref{phiaction0}. We focus on the diagram topology shown in Fig.~\ref{fig:topologyG1G2}(c), as it represents the lowest-order contribution in $G$ involving an internal vertex. As we will see, these contributions appear only  beyond the 3PN level and thus are not relevant for the results presented in this paper. We will justify this by analizing the PN order at which these effects begin to contribute.

The Gauss-Bonnet coupling introduces additional scalar-graviton vertices, beginning with a three-scalar-graviton-graviton vertex of the form $\sim \Phi h h$. Consequently, the lowest-order diagrams for the topology in Fig.~\ref{fig:topologyG1G2}(c) arise from wordline couplings to the scalar and gravitational fields that are linear in the spin. These couplings include $\alpha^0_{\rm a} m^0_{\rm a} \Phi$, $m^0_{\rm a} \phi$, and the leading spin coupling $S^{ij}_{\rm a} \partial_b A_a$, all derived from the point-particle Routhian \eqref{eq:mincouplingresults}. In this case, the three vertex stemming from the Gauss-Bonnet term contains necessarily the structure $\Phi \partial^2 \phi \partial^2 A_i$, where the indices are fully contracted and the partial derivatives consist of three spatial derivatives and one time derivative, {\it e.g.}, $\Phi \partial^i \dot{\phi} \partial^j\partial_i A_j$. Hence, using power counting, we can express the scaling of this contribution as follows:
\begin{equation}
	V \propto \alpha f'(\vphi_0) \frac{G^2 m^0_{\rm a} m^0_{\rm b}}{r^5} \alpha^0_{\rm a} S_{ij}^{\rm c} v^i_{\rm b}n^j\,,
\end{equation}
where ${\rm a}, {\rm b}$, and ${\rm c}$ can be either 1 or 2, depending on which wordline the vertices are attached to the diagram. For example, the coupling involving the scalar field can be attached to the worldline of particle 1, while the other two couplings are attached to the worldline of particle 2. Although there are other possibilities for this topology, they all exhibit the same scaling as above. For the discussion that follows, it is convenient to rewrite this contribution as
\begin{equation}\label{eq:scalingesgbvertex}
	V \propto \frac{\alpha f'(\vphi_0)}{(GM^0)^2} \left( \frac{G M^0}{r} \right)^2 \frac{G^2 m^0_{\rm a} m^0_{\rm b}}{r^3} \alpha^0_{\rm a} S_{ij}^{\rm c} v^i_{\rm b}n^j\,,
\end{equation}
where $M^0 = m_1^0 + m_2^0$ represents the total asymptotic mass of the system.

As discussed in \cite{Almeida:2024uph}, the inclusion of the Gauss-Bonnet term introduces a new length scale into the problem, which scales independently within the PN approximation. Nevertheless, in the case of ESGB gravity, one can still employ the ``small-$\alpha$" approximation, which assumes $\alpha f'(\vphi_0) \lesssim (G M^0)^2$, to estimate the equivalent PN order at which a particular term will contribute. This approximation is supported by observational constrains from recent GW events, which places $\sqrt{\alpha} \lesssim 1.7$km \cite{Julie:2024fwy,Nair:2019iur,Perkins:2021mhb,Wang:2021jfc}. Similar theoretical constraints, particularly for the shift-symmetric case, are discussed in Ref.~\cite{Sotiriou_2014b}. Using this information, it is easy to determine that the contribution \eqref{eq:scalingesgbvertex} scales as a 4.5PN term, which is well beyond the PN level considered in this paper.

\subsection{Final result}

To complete the two-body effective potential for spinning BHs in ESGB gravity up to the 3PN level, we need to incorporate several components. In addition to the pure GR results for spin contributions - with $m_{\rm a}$ replaced by $m^0_{\rm a}$ - and the non-spinning results for ESGB gravity, we must also include the curvature corrections coming from the worldline coupling in the Routhian \eqref{eq:routhianESGB}. 
It is important to note that these curvature corrections already contribute at the 3PN order via the $\mathcal{O}(G)$ diagram topology shown in Fig.~\ref{fig:topologyG1G2}(a), which necessarily involves a graviton mode. This situation excludes the possibility of new scalar interations at this level, meaning no additional contributions arise from scalar modes. Therefore, we can simply borrow the GR result and only replace $m_{\rm a}$ by $m^0_{\rm a}$. 
Additionally, as mentioned in Sec.~\ref{subsec:fsgravcoupling}, for the spin-induced quadrupole moment, we need to substitute $C^{({\rm a})}_{ES^2}$ in the GR results with $C^{(0)}_{ES^2,{\rm a}}$, as given by Eq.~\eqref{eq:gravcouplingmatching} evaluated at $\vphi = \vphi_0$. 

Thus, the final result for the effective two-body potential for spinning BHs in ESGB gravity, up to 3PN order and with spins in the covariant SSC, can be expressed as:
\begin{equation}\label{eq:finalresultforV}
	V = V^{\rm 3PN}_{\rm ESGB} + V^{\rm 3PN}_{\rm spin-GR} + V^{\rm 3PN}_0 + V^{\rm 3PN}_{h,{\rm fs}} + V^{\rm 3PN}_{\vphi,{\rm fs}}\,.
\end{equation}
In this expression, $V^{\rm 3PN}_{\rm ESGB}$ represents the non-spinning sector part of the effective potential at the 3PN order obtained in \cite{Julie:2022qux};  
the term $V^{\rm 3PN}_{\rm spin-GR}$ represents the spin sector of GR, with $m_{\rm a}$ replaced by $m^0_{\rm a}$ and $C^{(\rm a)}_{ES^2}$ by $C^{(0)}_{ES^2,{\rm a}}$, and further decomposed into:
\begin{equation}
	V^{\rm 3PN}_{\rm spin-GR} = V_{\rm SO} + V_{\rm SS} + V_{\rm S^2}\,,
\end{equation}
where $V_{\rm SO}$ is the spin-orbit part, which contains contributions at the 1.5PN and 2.5PN orders, computed in \cite{Porto:2010tr}; the term $V_{\rm SS}$
accounts for spin(1)spin(2) effects, which contribute at the 2PN and 3PN orders, obtained in \cite{rothstein2008spin1spin2}; The term $V_{\rm S^2}$ captures spin(1)spin(1) effects, which, in particular, contain contributions from the curvature corrections in the Routhian, as well as finite-size corrections, both of which computed in \cite{rothstein2008spin1spin1}. This last term also contains 2PN and 3PN parts; Finally, the remaining three pieces in \eqref{eq:finalresultforV} are computed for the first time in this paper and are given by
\begin{widetext}
\begin{align}
V_{0}^{\rm 3PN} &= \frac{G^2 m^0_2 \alpha^0_2}{r^3} n^i \left[ m^0_1 \alpha^0_1   \left( S_1^{0i} - S_1^{ij} (v_1^j - 2 v_2^j) \right) -\frac{1}{2} m^0_2 \alpha^0_2 S_1^{ij} (v_1^j - v_2^j) \right]  + (1 \leftrightarrow 2) \,, \\
V_{h,{\rm fs}}^{\rm 3PN} &= -\frac{G^2 m^0_2 \alpha^0_2}{2r^4}  \left( \alpha^0_1 C^{(0)}_{ES^2,1} + \frac{m^0_2}{m^0_1}  C^{(1)}_{ES^2,1}    \right)  S^{ik}_1 S^{jk}_1 (\delta_{ij} - 3n_i n_j) + (1\leftrightarrow 2)\,, \\
V_{\vphi,{\rm fs}}^{\rm 3PN} &= \frac{G m^0_2 \alpha^0_2}{2m^0_1 r^3} C^{(0)}_{\vphi S^2,1} \left[ \left( 1- \frac{v_1^2}{2}- \frac{v_2^2}{2} \right) S^{ik}_1 S^{jk}_1 
- ( S^{i0}_1 S^{j0}_1 + 2 S^{0k}_1 S^{ik}_1 v^j_2) \right] (\delta_{ij} - 3 n_i n_j) \nonumber \\
&~~~~ + \frac{G m^0_2 \alpha^0_2}{m^0_1 r^3} C^{(0)}_{\vphi S^2,1} S^{ik}_1 S^{jk}_1  v^i_1 [ v^j_1 - 3(\V_1\cdot \RR) n^j ] \nonumber\\
&~~~~ - \frac{ G^2 (m^0_2)^2 \alpha^0_2 }{ m^0_1 r^4 }
\left( 2   C^{(0)}_{\vphi S^2,1} + \frac{ \alpha^0_2 }{2}  C^{(1)}_{\vphi S^2,1}   \right)  S^{ik}_1 S^{jk}_1  (\delta_{ij} - 3 n_i n_j)
 + (1\leftrightarrow 2)\,,
\end{align}
\end{widetext}
where $V_{0}^{\rm 3PN}$ provides additional 2.5PN corrections to the effective potential, $V_{h,{\rm fs}}^{\rm 3PN}$ contributes with  3PN terms, and $V_{\vphi,{\rm fs}}^{\rm 3PN}$ with terms at both 2PN and 3PN orders.

One should keep in mind that, when rotation is included,  the above results are not the only new elements that are needed beyond the existing ESGB and GR results, to comprehensively characterize the dynamics of BH binaries. Rotation also modifies the BH sensitivities in the effective two-body potential by correcting the results of the non-spinning case, as discussed in Sec.~\ref{sec:thematching}. These parameters are crucial for an accurate effective description since they directly enter the effective potential. Because of this, we conclude this section with a brief analysis of how spins modify the sensitivity parameters.

\subsection{BH Sensitivities}

In a binary system, the sensitivities of a BH accounts for the adiabatic response of its ADM mass to the scalar field environment sourced by the companion. To study how these quantities are modified with the inclusion of spins, we use the analytical results for the mass function obtained in Sec.~\ref{sec:thematching}. Recall that the leading sensitivity $\alpha^0_{\rm a}$ is defined through a logarithmic derivative of $m_{\rm a}(\vphi)$ evaluated at the asymptotic value of the  scalar field $\vphi_0$:
\begin{equation}
\alpha^0_{\rm a} = \frac{d\log m_{\rm a}(\vphi)}{d\vphi}\bigg|_{\vphi\rightarrow\vphi_0}\,.
\end{equation}
Since the other subleading sensitivities $\beta^0_{\rm a}$, $\gamma^0_{\rm a}$, etc, are obtained from simple derivation of the above expression, $\alpha^0_{\rm a}$ plays a central role in this discussion. Because of this, we shall focus our analysis on this parameter, which is given in a generic ESGB theory by
\begin{equation}\label{eq:sensitgeneric}
\alpha^0_{\rm a} = \sum_{n=1}^7 \sum_{l=0,2,4} x^n z^l A^{(n)}_{l}(\vphi_0)\,,
\end{equation}
where
\begin{equation}
x \equiv \frac{\alpha f'(\vphi_0)}{\mu^2_{\rm a}} \qquad \text{and} \qquad z \equiv \frac{S_{\rm a}}{\mu^2_{\rm a}}\,.
\end{equation}
The two quantities $\mu_{\rm a}$ and $S_{\rm a}$, which as we have seen have a clear physical interpretation, fully characterize the sensitivities of the BHs. Up to $\mathcal{O}(x^4)$, the functions $A^{(n)}_l(\vphi_0)$ appearing in Eq.~\eqref{eq:sensitgeneric} are given by
\begin{widetext}
\begin{align}
A^{(1)}_0 &= - \frac12\,, \\
A^{(2)}_0 &= -\frac{f_0}{2 f'_0}-\frac{73 f''_0}{480 f'_0}\,, \\
A^{(3)}_0 &= -\frac{73}{480}  -\frac{f_0^2}{2 f_0'^2} -\frac{73 f_0 f_0''}{240 f_0'^2}-\frac{12511 f_0''^2}{241920 f_0'^2} -\frac{12511 f^{(3)}_0}{483840 f'_0}\,, \\
A^{(4)}_0 &= -\frac{f_0^3}{2 f_0'^3}-\frac{73 f_0}{160 f'_0}-\frac{799607 f^{(4)}_0}{255467520 f'_0}-\frac{12511 f^{(3)}_0 f_0}{161280 f_0'^2}-\frac{73 f_0^2 f''_0}{160f_0'^3} \nonumber\\
&-\frac{12511 f_0 f_0''^2}{80640 f_0'^3} -\frac{227192473 f_0''^3}{12773376000 f_0'^3}-\frac{5505779 f''_0}{26611200 f'_0}-\frac{31557593 f^{(3)}_0 f''_0}{1161216000 f_0'^2}\,,
\end{align}
for $l = 0$,
\begin{align}
A^{(1)}_2 &= \frac{1}{8}\,, \\
A^{(2)}_2 &= \frac{f_0}{2 f'_0}+\frac{21 f''_0}{320 f'_0}\,, \\
A^{(3)}_2 &= \frac{65}{384} +\frac{9 f_0^2}{8 f_0'^2}+\frac{65 f_0 f''_0}{192 f_0'^2}+\frac{3206149 f_0''^2}{106444800 f_0'^2} +\frac{3206149 f^{(3)}_0}{212889600 f'_0} \,, \\
A^{(4)}_2 &= \frac{2 f_0^3}{f_0'^3}+\frac{461 f_0}{480 f'_0} +\frac{461 f_0^2 f''_0}{480 f_0'^3} +\frac{806375 f^{(3)}_0 f_0}{8515584 f_0'^2} +\frac{148901437 f^{(4)}_0}{66421555200 f'_0} \nonumber\\
&+\frac{806375 f_0 f_0''^2}{4257792 f_0'^3}+\frac{4301149469 f_0''^3}{332107776000 f_0'^3}+\frac{56244203 f''_0}{251596800 f'_0}+\frac{58345277 f^{(3)}_0 f''_0}{2965248000 f_0'^2}\,,
\end{align}
for $l = 2$, and finally
\begin{align}
A^{(1)}_4 &= \frac{1}{16} \,, \\
A^{(2)}_4 &= \frac{9 f_0}{32 f'_0}+\frac{20687 f''_0}{1612800 f'_0} \,, \\
A^{(3)}_4 &= \frac{13267}{537600} + \frac{21 f_0^2}{32 f_0'^2} +\frac{13267 f_0 f''_0}{268800 f_0'^2}-\frac{60900137 f_0''^2}{17297280000 f_0'^2} -\frac{60900137 f_0^{(3)}}{34594560000 f'_0} \,,\\
A^{(4)}_4 &= \frac{9 f_0^3}{8 f_0'^3}+\frac{1397 f_0}{76800 f_0'} +\frac{1397 f_0^2 f''_0}{76800 f_0'^3} -\frac{1978930099 f^{(3)}_0 f_0}{92252160000 f_0'^2} -\frac{91508684921 f^{(4)}_0}{92990177280000 f'_0} \nonumber \\
& -\frac{1978930099 f_0 f_0''^2}{46126080000 f_0'^3} -\frac{196393841309 f_0''^3}{41094144000000 f_0'^3}-\frac{6898591302623 f_0''}{113008896000000 f'_0} -\frac{41938006913963 f^{(3)}_0 f''_0}{5424427008000000 f_0'^2} \,,
\end{align}
for $l = 4$. 
\end{widetext}

The generic expressions for the sensitivities of non-rotating BHs in generic ESGB gravity obtained in Ref.~\cite{BertiJulie2019} are recovered here with the above expressions for the $l=0$ functions $A^{(n)}_0$, with $n = 1,2,3,4$. The results for $l=2$ and $l=4$ are new in this paper and represent spin corrections to the non-rotating case in generic ESGB gravity. 
In the above expressions, $f_0, f'_0, \dots, f^{(k)}_0$ represent derivatives of $f(\vphi)$ evaluated at $\vphi_0$. The expressions for $A^{(n)}_l$ with $n=5,6,7$ are much bigger, and because of this, are available only in the ancillary file. 

It is instructive to look at a few of examples that are of particular interest in the literature, and explictly identify the regimes of $\alpha^0_{\rm a}$ where the binary dynamics significantly depart from GR. Let us first consider the cases of the dilatonic theory $f(\varphi)\propto e^{2\varphi}$ \cite{Kanti:1995vq} and the shift-symmetric theory $f(\varphi)\propto \varphi$ \cite{Yunes:2011we, Sotiriou_2014, Sotiriou_2014b}. The case of $f(\varphi)\propto \varphi$ is particularly interesting from an EFT point of view, as the $\varphi \mathcal{G}$ coupling is essentially the leading scalar-curvature correction that gives rise to hairy BHs even in the absence of the shift-symmetry \cite{Sotiriou_2014}.  As we will see, in both cases, the role of rotation in the sensitivity $\alpha^0_{\rm a}(\vphi_0)$ is mild, and, in particular, it does not change the general behavior of $\alpha^0_{\rm a}$ as a function of $\vphi_0$, but just delays the location of the pole, as predicted in Ref.~\cite{BertiJulie2019}. The same also happens for the other cases of interest, namely the quartic and Gaussian theories, which are discussed in 
Appendix~\ref{sec:sensSSandGauss}. The choice for the latter two cases lies in the context of spontaneous scalarization of BHs \cite{Doneva:2017bvd, Silva:2017uqg, Doneva:2022ewd}, where the BH scalarizes according to its curvatures. Also, complete expressions for the sensitivities in the four cases are provided in the ancillary Mathematica notebook.

\subsubsection{Dilatonic theory}

The dilatonic theory is defined by the choice of 
\begin{equation}
f(\vphi) = \frac{e^{2\vphi}}{4} \,,
\end{equation}
which makes the action \eqref{phiaction0} invariant under the simultaneous redefinition
\begin{equation}\label{eq:redefEDGB}
\vphi \rightarrow \vphi + c \qquad \text{and} \qquad \alpha \rightarrow \alpha e^{-2c}\,,
\end{equation}
where $c$ is a constant.
From Eq.~\eqref{eq:sensitgeneric}, the sensitivity in this case (shown here up to $\mathcal{O}(x^4,z^4)$) is given by
\begin{align}
\alpha^0_{\rm a}&=-\frac{x}{2} -\frac{133 x^2}{240} -\frac{35947 x^3}{40320} -\frac{474404471 x^4}{266112000} \nonumber\\
& +\left( \frac{x}{8}+\frac{61 x^2}{160} +\frac{17204749 x^3}{17740800} +\frac{17329330303x^4}{6918912000} \right) z^2  \nonumber\\
&+\bigg( \frac{x}{16} +\frac{134087 x^2}{806400} +\frac{1251010451 x^3}{5765760000} \nonumber\\
&-\frac{14525228147627 x^4}{72648576000000} \bigg) z^4 \,,
\end{align}
with
\begin{equation}
x = \frac{\alpha e^{2\vphi_0}}{2 \mu^2_{\rm a}}\,.
\end{equation}
This analytic expression generalizes to the rotating case the BH sensitivity in  Einstein-dilaton-Gauss-Bonnet theory studied in Sec.~IV B of Ref.~\cite{BertiJulie2019}. Note also that, this sensitivity is preserved under the redifinition \eqref{eq:redefEDGB}.

\begin{figure}
\centering
	\includegraphics[scale=0.63]{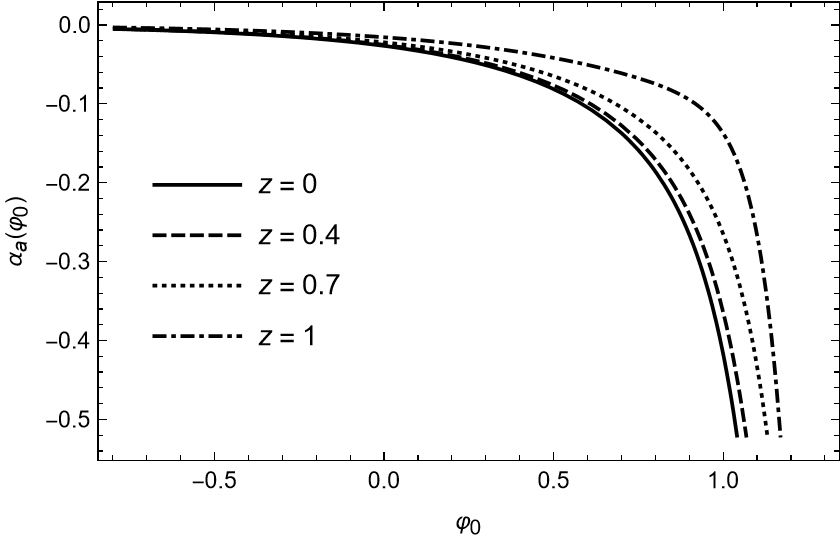}
\caption{Effects of rotation on the BH sensitivity in the dilatonic theory (with $f(\vphi) = {e^{2\vphi}}/{4}$). Plotted are $\alpha_{\rm a} = \alpha_{\rm a}(\vphi_0)$ for a BH with $x = \alpha/\mu_{\rm a} = 0.1$ and various values of the rotation parameter $z = S_{\rm a}/\mu_{\rm a}^2$, specifically $z=0, 0.4, 0.7, 1$. The general behavior of these curves remains the same for different values of $x$.}
\label{fig:sensitivities}
\end{figure}

To investigate how rotation affects the behavior of the sensitivity in this case, we plot in Fig.~\ref{fig:sensitivities}  $\alpha^0_{\rm a} = \alpha^0_{\rm a}(\vphi_0)$ in terms of $\vphi_0$ for a fixed value of $x$ and some representative values of $z$. In these plots, we use our most accurate expression for $\alpha^0_{\rm a}$ available, {\it i.e.}, at $\mathcal{O}(x^7,z^4)$. Compared to the non-rotating case, we see that the general behavior of this quantity is unchanged. Namely, from Fig.~\ref{fig:sensitivities}, we see that:
\begin{enumerate}
	\item The BH decouples from the scalar field as $\vphi_0 \rightarrow - \infty$, as a result of $\alpha^0_{\rm a} \rightarrow 0$ and $\beta^0_{\rm a} \rightarrow 0$;
	\item The BH couples more strongly to the scalar field as $\vphi_0$ increases.
\end{enumerate}
Additionally, we see that rotation delays the plunging of the curve for a given value of $\mu_{\rm a}$. The rotating case departs more and more from the non-rotating curve (solid line in Fig.~\ref{fig:sensitivities}) as the BH's spin $S_{\rm a}$ increases.

Following the discussions of Ref.~\cite{BertiJulie2019}, we use a Pad\'{e} approximant for $\alpha^0_{\rm a}$ to predict the new value of the pole $\vphi^{\rm pole}_0$. This is already known in the non-rotating case, whose existence has been confirmed numerically \cite{Julie:2022huo}. We investigate how it gets modified with the introduction of rotation. The Pad\'{e} resummation is a well-known technique to improve the convergence of the series, incorporating some non-perturbative effects. Indeed, in Ref.~\cite{BertiJulie2019} this technique was used to predict the existence of a pole in $\alpha^0_{\rm a}$. The origin of this pole can be traced back to bounds on the allowed values for the scalar field at the event horizon of a non-rotating BH, $\vphi_{\rm h}$, obtained non-perturbatively and given in generic ESGB gravity by \cite{Doneva:2017bvd}
\begin{equation}\label{eq:ineqgenericnonrot}
24 \alpha^2 f'(\vphi_{\rm h})^2 < (A_{\rm h}/4\pi)^2\,,
\end{equation}
where $A_{\rm h}$ is the area of the BH's event horizon. Because of this bound, the sensitivities will naturally become singular as $\vphi_0$ increases and the scalar field at the horizon approaches the critical value limited by Eq.~\eqref{eq:ineqgenericnonrot}. In the dilatonic theory, the inequality \eqref{eq:ineqgenericnonrot} reduces to
\begin{equation}\label{eq:ineqvphih}
\frac{\alpha e^{2\vphi_{\rm h}}}{2\mu^2_{\rm a}} < \frac{2}{1+\sqrt{6}}\,.
\end{equation}
Then, using the diagonal (2,2) Pad\'{e} approximant for the sensitivity $\alpha^0_{\rm a}$, represented by
\begin{equation}
\alpha^0_{\rm a,\text{Pad\'{e}}} = \mathcal{P}^2{}_2 [\alpha^0_{\rm a},x]\,,
\end{equation}
and keeping corrections up to $\mathcal{O}(z^4)$, we obtain
\begin{equation}
x_{\rm pole} = \frac{\alpha e^{2\vphi^{\rm pole}_0}}{2\mu^2_{\rm a}} = 0.4455 +0.1126 z^2 -0.0652 z^4 \,,
\end{equation}
which generalizes Eq.~(IV.17) of Ref.~\cite{BertiJulie2019}. From this, we see that the value of the pole is increased with $S_{\rm a}$, as compared to its non-rotating counterpart.

Note that the above result seems to indicate that the upper limit of the inequality \eqref{eq:ineqvphih} will possibly increase, allowing for the existence of BHs with a large value of $\vphi_{\rm h}$ as compared to the non-rotating case. A numerical study is necessary to confirm this prediction and to further exploit the effect of rotation on the sensitivities, where the rotation parameter can be consistently extrapolated up to the extremal case.

\subsubsection{Shift-symmetric theory}

The shift-symmetric theory is defined by
\begin{equation}
f(\vphi) = 2\vphi \,.
\end{equation}
The action for this theory is invariant under $\vphi \rightarrow \vphi + const$. As mentioned, what matters for the following results is the form of $f(\vphi) = 2 \vphi$, not the shift-symmetry {\it per se}. From an EFT point of view, the linear coupling $\varphi \mathcal{G}$ is the leading term in the expansion of $f(\vphi)$, regardless of the shift-symmetry.
From Eq.~\eqref{eq:sensitgeneric}, the sensitivity in this case (displayed up to $\mathcal{O}(x^4,z^4)$) is given by
\begin{widetext}
\begin{align}
\alpha^0_{\rm a} &= -\frac{x}{2} 
-\frac{\varphi_0  x^2}{2} 
- \left( \frac{73}{480} + \frac{\varphi_0 ^2}{2} \right) x^3
- \left( \frac{73 \varphi_0 }{160} +\frac{1}{2} \varphi_0^3 \right) x^4
+\left[ \frac{x}{8} +\frac{\varphi_0  x^2}{2} + \left( \frac{65}{384} +\frac{9 \varphi_0^2}{8} \right) x^3 + \left( \frac{461 \varphi_0}{480} +2 \varphi_0 ^3 \right) x^4 \right] z^2  \nonumber\\
&+ \left[
\frac{x}{16} + \frac{9 \varphi_0  x^2}{32} 
+\left( \frac{13267 x^3}{537600}
+\frac{21 \varphi_0^2 x^3}{32} \right)
+\left( \frac{1397 \varphi_0 x^4}{76800}
+\frac{9 \varphi_0 ^3 x^4}{8} \right)
\right] z^4 \,,
\end{align}
\end{widetext}
with
\begin{equation}
x = \frac{2 \alpha}{\mu^2_{\rm a}}\,.
\end{equation}
This expression generalizes to the rotating case the sensitivity for the shift-symmetric theory studied in Appendix E($b$) of Ref.~\cite{BertiJulie2019}.

Using the diagonal (2,2) Pad\'{e} approximant and keeping just the leading correction with the spin, we obtain
\begin{align}
\vphi^{\rm pole}_0 &= \frac{1}{2} \left( \frac{\mu^2_{\rm a}}{\alpha} - \frac{\sqrt{1095}}{30} \right)  \nonumber\\
&+ \left( \frac{ \sqrt{1095}\mu^4_{\rm a} }{292 \alpha^2} + \frac{179\mu^2_{\rm a}}{584 \alpha} - \frac{21\sqrt{1095}}{2920} \right) z^2 \,,
\end{align}
which extends Eq.~(E.10) of Ref.~\cite{BertiJulie2019} to the rotating case. Hence, similarly to the dilaton case, for $\mu^2_{\rm a} / \alpha > 0.630$, the value of the pole is increased with $S_{\rm a}$ with respect to its counter part in the non-rotating case. 

The behavior of $\alpha_{\rm a}^0$ in this case is similar to that of Fig.~\ref{fig:sensitivities}) for the dilaton case, analogous to those of the non-rotating BHs \cite{BertiJulie2019}. Also, like in the previous case, the role of the rotation parameter $z$ is to parametrize small deviations from the curve of vanishing spin, $z=0$.

\section{Conclusions}\label{sec:conclusion}

In this work, we have investigated the PN dynamics of rotating BHs within the framework of generic ESGB gravity theories. To achieve this, we extended the EFT framework for spins in GR, which makes use of a Routhian formalism, to incorporate the new scalar degree of freedom. This extension then allowed us to derive the complete conservative dynamics of BH binary systems in ESGB gravity to the 3PN accuracy, including spin corrections. To connect the effective field description to the single BH solutions, we performed a matching procedure, employing spinning BH solutions obtained in this paper 
perturbatively up to order $\mathcal{O}(\epsilon^7,\chi^5)$. In this process, we were able to derive the sensitivities of spinning BHs, which are crucial coefficients entering the effective two-body potential. These coefficients, shown to be related to the intrinsic thermodynamical properties of the BHs, are important for accurately describing the dynamics of inspiraling BH binaries in ESGB gravity. Therefore, we also examined the thermodynamical properties of our solutions for their physical relevance in this problem.

The results obtained in this paper extend previous work in several ways. First, spin effects had not been considered previously in the PN dynamics of BH binaries in ESGB gravity. Not only did we address this gap, but we also computed spin corrections up to the 3PN order, representing the most advanced computation of the conservative sector for binary motion in this class of theories. Additionally, to the best of our knowledge, spin corrections had not been explored in the study of sensitivities before. Moreover, the isolated BH solutions obtained here extend to generic ESGB theories the dilatonic theory found in Ref.~\cite{Maselli:2015tta}. The choice of $\mathcal{O}(\epsilon^7,\chi^5)$ in this work is based on the latter reference, where errors of at most $1\%$ for quantities of astrophysical relevance are achieved when compared to numerical studies. 

Possible future directions include conducting a detailed numerical study of the sensitivities of spinning BHs in ESGB gravity, similar to the approach taken in Ref.~\cite{Julie:2022huo} for the non-rotating case. Such a study could capture non-perturbative effects not covered by our analytic results, particularly in models where BHs may undergo spin-induced scalarization \cite{Herdeiro:2020wei, Berti:2020kgk}. Additionally, the highly accurate BH solutions obtained in this paper could be leveraged for analytical investigations into the astrophysical aspects of these solutions, especially those subject to spin-induced scalarization. Furthermore, our PN results could be utilized to develop GW templates for spinning binaries, as done in Ref.~\cite{Julie:2024fwy} for non-rotating binaries in ESGB gravity. In that work, the authors constructed complete inspiral-merger-ringdown waveforms using the effective-one-body formalism and the previously available PN results for non-rotating binaries at the 3PN level.

\section*{Acknowledgments}

We would like to thank Wen-Kai Nie and Jun Zhang for helpful discussions. SYZ acknowledges support from the National Key R\&D Program of China under grant No.~2022YFC2204603 and from the National Natural Science Foundation of China under grant No.~12075233. GLA and SYZ also acknowledge support from the National Natural Science Foundation of China under Grant No.~12247103.

\appendix

\section{Spinning bodies in NRGR}\label{sec:spinGRreview}

The EFT extension for spinning objects within the PN framework was originally devised by Porto \cite{Porto:2005ac}. In this approach, additional degrees of freedom accounting for rotation are introduced at the level of the worldline effective action, and encoded by the local Lorentz matrix $\Lambda^A{}_a(\lambda)$ that transforms the locally flat frame $e^a_\mu$ (satisfying $g^{\mu\nu}e^a_{\mu} e^b_\nu = \eta^{ab}$) to the co-rotating frame $e^A_\mu$, {\it i.e.}, $e^A_\mu = \Lambda^A{}_a(\lambda) e^a_\mu$, which is defined along the particle's trajectory. The generalized angular velocity is defined in terms of the co-rotating basis by $\Omega^{\mu\nu} = \eta^{AB}e^\mu_A\frac{De^\nu_B}{d\lambda}$, where $\frac{D}{d\lambda} = \frac{dx^\mu}{d\lambda} \nabla_\mu$, and the antisymmetric spin tensor $S^{\mu\nu}$ defined as the conjugate momentum to $\Omega_{\mu\nu}$.

As originally investigated in \cite{Porto:2005ac}, which extends the work of Hanson and Regge beyond the realm of special relativity \cite{hanson1974}, a minimal effective action can be fixed by general covariance and invariance under worldline reparametrization, taking the form:
\begin{equation}\label{eq:ppactionspin}
S =  \sum_{\rm a} \int d\lambda_{\rm a} \left( p^\mu_{\rm a} u^{\rm a}_\mu + \frac12 S_{\rm a}^{\mu\nu}\Omega^{\rm a}_{\mu\nu} \right)\,,
\end{equation}
from which the Mathisson-Papapetrou-Dixon (MPD) equations follow from the variational principle \cite{Porto:2005ac} (see also Appendix A of \cite{armaza2016} for an alternative derivation):
\begin{equation}\label{eq:MPDeqs}
    \frac{Dp^\mu_{\rm a}}{d\lambda} = -\frac12 R^\mu{}_{\nu\alpha\beta} u^\nu_{\rm a} S^{\alpha\beta}_{\rm a}\,, ~~\frac{DS^{\mu\nu}_{\rm a}}{d\lambda} = p^\mu_{\rm a} u^\nu_{\rm a} - p^\nu_{\rm a} u^\mu_{\rm a}\,.
\end{equation}
As is well known, this set of equations governs the dynamics of spinning particles on a generic curved background \cite{mathisson1940a,mathisson1940b}. Here, $p^\mu_{\rm a}$ represent the particles' four-momenta, which now depend on the spin, differing from the usual expression $p^\mu_{\rm a} = m u^\mu_{\rm a}$ (if $\lambda_{\rm a}$ is the proper time).

By rewriting the action \eqref{eq:ppactionspin} in terms of the locally flat frame $e^a_\mu$, the spin coupling to gravity can be factored out and given in terms of the Ricci rotation coefficients $\omega^{ab}_\mu = e^b_\nu \nabla_\mu e^{a\nu}$: 
\begin{equation}
S_{\rm spin-gravity} = \frac12 \sum_{\rm a} \int d\lambda_{\rm a}\, S^{\rm a}_{ab} \omega_\mu^{ab} u^\mu_{\rm a}\,,
\end{equation}
where $S^{ab}\equiv S^{\mu\nu}e^a_{\mu}e^b_{\nu}$ is the spin tensor in the locally flat frame. From this coupling, Feynman rules can be derived, allowing the computation of the effect of spins on the two-body potential.

Following the above discussion, a spin supplementary condition (SSC) must be imposed to eliminate the unphysical degrees of freedom of the spin tensor from physical observables. In the initial studies of spins in NRGR \cite{Porto:2005ac,rothstein2006prl}, the so-called Newton-Wigner SSC was used and imposed directly at the Lagrangian level. While this route has the advantage of allowing us to work directly with the three physical degrees of freedom of the spin tensor from the start, it has the drawback of 
complicating the canonical structure of the Possion brackets \cite{Porto:2005ac,hanson1974}. To circumvent this, a new approach was put forward in Ref.~\cite{porto2007proceedings} in the context of the covariant SSC by making use of the Routhian formalism. In this approach, the preservation of the SSC upon evolution is used to constrain the new action, which then yields the introduction of new terms to the worldline effective action, but it does so without altering the canonical structure of the  Poisson brackets. We will make use of this formalism for ESGB gravity, and briefly review it in the following.

\subsection{The Routhian for spinning bodies}

Given that the Lagrangian for spinning objects is a function of the angular velocity, $L(x^\mu,u^\mu,\Omega^{\mu\nu})$, without explicit dependence on the rotating degrees of freedom, we can conveniently perform a partial Legendre transformation on the angular variables. This results in a function $\mathcal{R}$ that behaves as a Lagrangian with respect to the positions and velocities while acting as a Hamiltonian with respect to the spin. This hybrid formalism is called Routhian, and the same name is given to the function obtained through the (partial) Legendre transformation. 

In this approach, we decompose the angular velocity into two components: a local part, given by the local angular velocity 
\begin{equation}
    \Omega_L^{ab} = \eta_{AB} \Lambda^A{}_a \frac{D\Lambda^B{}_b}{d\lambda}\,,
\end{equation}
and a gravitational part, expressed in terms of the Ricci rotation coefficients, written as $e^a_\mu e^b_\nu \Omega^{\mu\nu} = \Omega^{ab}_L + u^\mu \omega_\mu^{ab}$. This splitting allows us to separate the pure spin degrees of freedom from the spin-gravity coupling, and we only pass to the Routhian formulation for the local part
\begin{align}
\mathcal{R} &= L - \frac12 S_{ab}\Omega_L^{ab} \nonumber\\
&= p^\mu u_\mu  + \frac12 \omega_\mu^{ab} S_{ab} u^\mu\,. \label{eq:routhianPU}
\end{align}

The equations of motion for positions and spins are obtained from: 
\begin{equation}\label{eq:EOMs}
\frac{\delta}{\delta x^\mu} \int d\lambda\,\mathcal{R} = 0 \qquad \text{and} \qquad \frac{dS^{ab}}{d\lambda} = \{S^{ab},\mathcal{R}\}\,,
\end{equation}
with Poisson brackets given by the Lorentz algebra:
\begin{equation}\label{lorentzalgebra}
\{ S^{ab}, S^{cd}\} = \eta^{ad} S^{bc} + \eta^{bc} S^{ad} - \eta^{ac} S^{bd} - \eta^{bd} S^{ac} \,.
\end{equation}
In practice, we parametrize the worldlines by the ordinary time, $\lambda = t$, and set $\mathcal{R} = - V$ in \eqref{eq:EOMs}, since spin effects will enter through corrections to the effective two-body potential $V$.

We now address the issue of incorporating an SSC into the action. As previously mentioned, the key advantage of the Routhian formalism is its ability to consistently impose an SSC without altering the canonical structure of the Poisson brackets. In particular, here, we follow the discussion in \cite{rothstein2008spin1spin2} and consider the covariant SSC:
\begin{equation}\label{eq:SSCpS}
p_a S^{ab} = p_\mu S^{\mu\nu} = 0\,.
\end{equation}
The requirement that this condition be preserved upon dynamical evolution, {\it i.e.},
\begin{equation}
\frac{D}{d\lambda} (p_\mu S^{\mu\nu}) = 0\,,
\end{equation}
allows us to derive, with the help of the MPD equations, an expression for the momentum in terms of the four-velocity and curvature corrections:
\begin{equation}\label{eq:momentumPUR}
p^{\mu} = \frac{1}{\sqrt{-u^2}}  \left( m u^\mu + \frac{1}{2m} R_{\beta\nu\rho\sigma} S^{\mu \nu} S^{\rho\sigma} u^\beta \right) + \dots\,.
\end{equation}
The ellipses in this expression represent higher-order curvature terms that are irrelevant to our 3PN level, as they only contribute to higher orders in the effective two-body potential. Notice that, ignoring curvature corrections, we have that $m= -p\cdot u $ (for $\lambda$ the proper time). Hence, replacing \eqref{eq:momentumPUR} into \eqref{eq:routhianPU} we arrive at the final form of our Routhian:
\begin{align}
\mathcal{R} = \sum_{\rm a} \bigg( &- m_{\rm a} \sqrt{-u_{\rm a}^2} + \frac12 S^{\rm a}_{ab} \omega^{ab}_{\mu} u^\mu_{\rm a} \nonumber\\
&+ \frac{1}{2m_{\rm a}} R_{abcd} S^{ab}_{\rm a} S^{ce}_{\rm a} \frac{u^d_{\rm a} u^{\rm a}_e}{\sqrt{-u^2_{\rm a}}} \bigg) + \dots\,, \label{routhian}
\end{align}
where, again, the curvature corrections are needed to guarantee the preservation of the covariant SSC during the dynamical evolution of the system.

The Routhian \eqref{routhian} modifies the point-particle action in NRGR to incorporate spins under the covariant SSC. Hence, combined with the Einstein-Hilbert action and a gauge-fixing term, spin corrections to the effective two-body potential $V$ can be calculated, as outlined in Sec.~\ref{sec:EFTsetup}. In this case, after deriving the contributions to the equations of motion using \eqref{eq:EOMs}, the SSC condition \eqref{eq:SSCpS} must be applied to eliminate the $S^{0i}$ components. This ensures that the final results are expressed solely in terms of the three spin degrees of freedom of $S^{ij}$. Note that it is crucial that the SSC is not imposed at the Lagrangian level, as this would modify the canonical structure of the Poisson brackets.	

Finally, the last ingredient we need for the PN approximation is a power counting for the spin, which is given by $S^{ij} = [\text{moment of inertia}]\times[\text{angular velocity}] \sim m r_s^2 \times \frac{v_{\rm r}}{r_s}$, where $v_{\rm r}$ is the body's typical rotational speed. Hence, assuming the case of a maximally rotating limit, $v_{\rm r} \approx 1$, we obtain $S^{ij} \sim mrv^2$. In addition, because of the SSC, we have $S^{0i} \sim S^{ij} v^j$.

\subsection{Finite-size corrections}

In the non-spinning case, finite-size operators that account for tidal deformability effects become relevant only at the 5PN order. 
Nevertheless, when spins are considered, rotation-induced deviations from spherical symmetry require finite-size operators to be added to the worldline effective action, which then induces the contributions to the two-body effective potential to the 2PN order \cite{rothstein2008spin1spin1}. In this case, the physically relevant higher-dimensional operators are expressed in terms of the compact objects' rotation-induced multipole moments, which couple linearly to the electric, $E_{ab}$, and magnetic, $B_{ab}$, components of the Weyl tensor \cite{Porto:2005ac,rothstein2008spin1spin1}.

At the 3PN level of binary dynamics, the coupling of the quadrupole moment to $E_{ab}$ is the only operator of this kind that contributes, and is given by: 
\begin{equation}\label{eq:finitesizeSh}
S^h_{\rm fs} =  -\frac12 \int d\lambda\, Q^{ab}_E \frac{E_{ab}}{\sqrt{-u^2}}\, ,
\end{equation}
where the quadrupole moment of the compact object $Q^{ab}_E$ is defined by
\begin{equation}
Q^{ab}_E = \frac{C_{ES^2}}{m} S^a{}_c S^{bc}\,,
\end{equation}
and the electric component of the Weyl tensor $E_{ab}$ is: 
\begin{equation}
E_{ab} = C_{acbd} u^c u^d\,.
\end{equation}
The coefficient $C_{ES^2}$ can be obtained through a matching procedure. For the case of a rotating BH, {\it i.e.}, the Kerr BH in the GR case, $C_{ES^2} = 1$. For the case of neutron stars, the typical value ranges between $C_{ES^2}\sim 4-8$ depending on the equation of state \cite{poisson1997eosNSa,poisson1997eosNSb}.

\section{Sensitivities for specific theories}\label{sec:sensSSandGauss}

In this appendix, we present the results for the sensitivies of rotating BHs in ESGB gravity in two other cases of great relevance: the quartic and the Gaussian theories. These theories give rise to the interesting phenomenon of spontaneous scalarization.

\subsubsection{Quartic theory}

The quartic theory is defined by
\begin{equation}
    f(\vphi) = \frac{1}{8} \eta \vphi^2 + \frac{1}{16} \zeta  \phi^4.
\end{equation}
In this case, we have 
\begin{equation}
    x = \frac{\alpha \vphi_0 (\eta + \zeta \vphi_0^2)}{\mu_{\rm a}^2}\,.
\end{equation}
Then, defining $\kappa \equiv \eta + \zeta \vphi_0^2$\,, below we display the sensitivity $\alpha^0_{\rm a}$ up to $\mathcal{O}(x^4,z^2)$: 
\begin{widetext}
\begin{align}
\alpha^0_{\rm a} &=
-\frac{x}{2} - \left( \frac{73}{160 \vphi_0} + \frac{\eta \vphi_0}{8 \kappa } + \frac{\vphi_0 }{8}-\frac{73 \eta }{240 \kappa 
   \vphi_0} \right) x^2 \nonumber\\
& + \left( -\frac{73 }{192}  -\frac{\eta  \vphi_0^2}{16 \kappa }-\frac{\eta ^2 \vphi_0^2}{32 \kappa
   ^2}-\frac{\vphi_0^2 }{32}-\frac{73 \eta  }{960 \kappa }+\frac{73 \eta ^2 }{480 \kappa ^2}+\frac{12511 \eta }{16128 \kappa  \vphi_0^2}-\frac{12511 \eta ^2}{60480 \kappa ^2
   \vphi_0^2}-\frac{12511}{20160 \vphi_0^2} \right) x^3 \nonumber\\   
& +\bigg(-\frac{3 \eta ^2 \vphi_0^3}{128 \kappa ^2}-\frac{\eta ^3 \vphi_0^3 }{128 \kappa ^3}-\frac{\vphi_0^3 }{128}-\frac{73
   \eta  \vphi_0 }{320 \kappa }+\frac{73 \eta ^2 \vphi_0 }{2560 \kappa ^2}+\frac{73 \eta ^3 \vphi_0 }{1280 \kappa ^3}-\frac{511 \vphi_0
   }{2560} +\frac{28216061 \eta  }{53222400 \kappa  \vphi_0} +\frac{137621 \eta ^2 }{322560 \kappa ^2 \vphi_0}\nonumber\\
&-\frac{3 \eta  \vphi_0^3 }{128 \kappa } -\frac{12511 \eta ^3 }{80640 \kappa ^3 \vphi_0}-\frac{9634409 }{8870400 \vphi_0}+\frac{636730037 \eta  }{354816000 \kappa  \vphi_0^3}-\frac{73479353 \eta ^2 }{76032000 \kappa ^2 \vphi_0^3} +\frac{227192473 \eta^3 }{1596672000 \kappa ^3 \vphi_0^3}-\frac{120214117 }{121651200 \vphi_0^3} \bigg) x^4 \nonumber \\
&+z^2 \bigg[ \frac{x}{8}
+ \left( \frac{\eta  \vphi_0 }{8 \kappa }+\frac{\vphi_0 }{8}-\frac{21 \eta }{160 \kappa  \vphi_0}+\frac{63}{320 \vphi_0} \right) x^2 
+\bigg( \frac{9 \eta  \vphi_0^2 }{64 \kappa }+\frac{9 \eta ^2 \vphi_0^2 }{128 \kappa ^2}+\frac{9 \vphi_0^2 }{128} +\frac{65 \eta }{768 \kappa }-\frac{65 \eta ^2}{384 \kappa ^2}-\frac{3206149 \eta}{7096320 \kappa  \vphi_0^2}
\nonumber\\
&+\frac{3206149 \eta ^2}{26611200 \kappa ^2 \vphi_0^2}+\frac{3206149 }{8870400 \vphi_0^2}+\frac{325}{768} \bigg) x^3 
+\bigg( \frac{3 \eta  \vphi_0^3}{32 \kappa }+\frac{3 \eta ^2 \vphi_0^3
   }{32 \kappa ^2}+\frac{\eta ^3 \vphi_0^3 }{32 \kappa ^3}+\frac{\vphi_0^3 }{32} +\frac{461 \eta  \vphi_0 }{960 \kappa }-\frac{461 \eta ^2 \vphi_0 }{7680
   \kappa ^2}-\frac{461 \eta ^3 \vphi_0 }{3840 \kappa ^3}
\nonumber\\
&+\frac{3227 \vphi_0 }{7680}-\frac{3260960557 \eta  }{5535129600 \kappa  \vphi_0} -\frac{806375 \eta ^2 }{1548288
   \kappa ^2 \vphi_0} +\frac{806375 \eta ^3 }{4257792 \kappa ^3 \vphi_0}+\frac{1142829983 }{922521600 \vphi_0}-\frac{12021367921 \eta  }{9225216000 \kappa  \vphi_0^3}
\nonumber\\   
&+\frac{19438119431 \eta ^2 }{27675648000 \kappa ^2 \vphi_0^3}-\frac{4301149469 \eta ^3 }{41513472000 \kappa ^3 \vphi_0^3}
+\frac{15881477147}{22140518400 \vphi_0^3} \bigg) x^4
   \bigg] \,. 
\end{align}
\end{widetext}
The complete expression valid up to $\mathcal{O}(x^7,z^4)$ is provided in the ancillary Mathematica notebook.

\begin{figure}
\centering
	\includegraphics[scale=0.63]{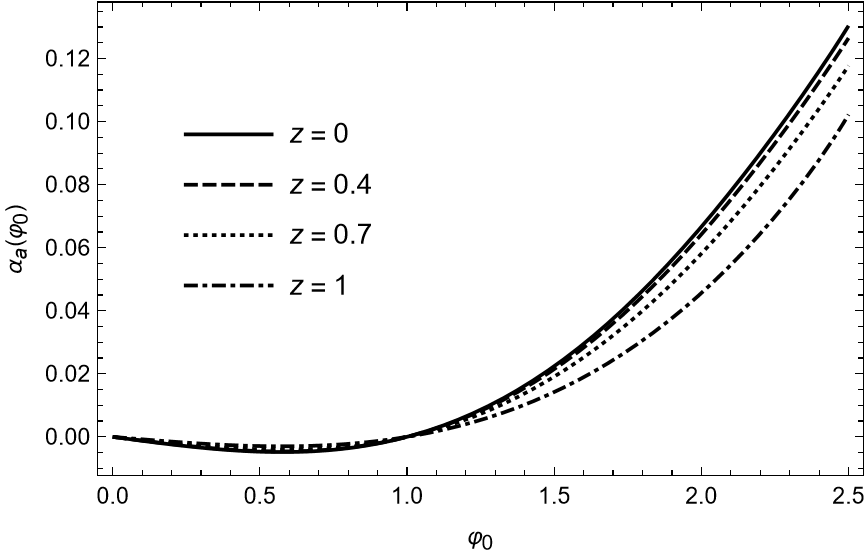}
\caption{Effects of rotation on the BH sensitivity in the quartic theory, $f(\vphi) = \frac{1}{8} \eta \vphi^2 + \frac{1}{16} \zeta  \phi^4$, with $\eta = 1$ and $\zeta = -1$. Plotted are $\alpha_{\rm a} = \alpha_{\rm a}(\vphi_0)$ for a BH with $x = \alpha/\mu_{\rm a} = 0.1$ and various values of the rotation parameter $z = S_{\rm a}/\mu_{\rm a}^2$, specifically $z=0, 0.4, 0.7, 1$. The general behavior of these curves remains the same for different values of $x$.}
\label{fig:sensitivities2}
\end{figure}

In Fig.~\ref{fig:sensitivities2}, we plot the sensitivity $\alpha_{\rm}(\vphi_0)$ for $x = 0.1$ and various values of the rotation parameter $z$ for the quartic theory with $\eta = 1$ and $\zeta = -1$. In this case, we see that the general behavior of the curves is similar to the dilatonic and shift-symmetric theories, but with opposite sign. Also, like in the others case, different values of $x$ do not change this behavior.

We emphasize here that, due to the importance of the quartic theory, as it provides an example of a theory that exhibits the phenomenon of spontaneous black hole scalarization, a numerical study can greatly benefit the analysis of the sensitivities in this case. In particular, one question one my address is whether the negative values of $\alpha_{\rm a}^0$ lying between $\vphi_0 = 0$ and $\vphi_0 = 1$ in the curves of Fig.~\ref{fig:sensitivities2} is simply an artifact of our perturbative solution, or if it carries any physical content. \\

\subsubsection{Gaussian theory}

The Gaussian theory is defined by
\begin{equation}
f(\vphi) = -\frac{1}{12} e^{-6\vphi^2}\,.
\end{equation}
In this case, the sensitivity $\alpha^0_{\rm a}$ is given by: (up to $\mathcal{O}(x^4,z^4)$)
\begin{widetext}
\begin{align}
\alpha^0_{\rm a} &= -\frac{x}{2} -\left(\frac{53}{480 \varphi_0 } - \frac{73 \varphi_0  }{40} \right) x^2
+\left(\frac{9883 }{5760}  -\frac{7219 }{241920 \varphi_0^2}  -\frac{12511 \varphi_0^2 }{1120}\right) x^3 \nonumber \\
&-\left( \frac{98822473 }{12773376000 \varphi_0 ^3} - \frac{80401577 }{76032000 \varphi_0 } + \frac{1842044651 \varphi_0  }{88704000} - \frac{102384391 \varphi_0 ^3 }{1232000} \right) x^4 \nonumber\\
&+ \bigg[ \frac{x}{8} + \left( \frac{23}{960 \varphi_0 } - \frac{63 \varphi_0  }{80} \right) x^2
+\left( \frac{1034749}{106444800 \varphi_0 ^2} -\frac{1919149}{2534400} +\frac{3206149 \varphi_0 ^2}{492800} \right) x^3 \nonumber\\
&+\left( \frac{890340769}{332107776000 \varphi_0 ^3} -\frac{2439852509}{4612608000 \varphi_0 } +\frac{29460157033 \varphi_0}{2306304000} -\frac{1930054613 \varphi_0 ^3}{32032000} \right) x^4
\bigg] z^2 \nonumber\\
&+ \bigg[
\frac{x}{16}
-\left(
\frac{17113 }{1612800 \varphi_0 } +\frac{20687 \varphi_0 }{134400} \right) x^2 -\left(  \frac{106431349}{34594560000 \varphi_0 ^2} -\frac{1279467643}{5765760000} +\frac{60900137 \varphi_0 ^2}{80080000} \right) x^3
\nonumber\\
&- \left( \frac{781398796049}{452035584000000 \varphi_0 ^3} -\frac{3409941808201}{12556544000000 \varphi_0 } +\frac{110433695435387 \varphi_0}{18834816000000} -\frac{54900000440357 \varphi_0 ^3}{2354352000000} \right) x^4 \bigg] z^4 \,,
\end{align}
\end{widetext}
with
\begin{equation}
x = \frac{\alpha \vphi_0 e^{-6\vphi_0^2}}{\mu^2_{\rm a}}\,.
\end{equation}

This sensitivity, which includes spin effects, behaves in the same way as analyzed in Ref.~\cite{BertiJulie2019}, with the values of $z$ parametrizing mild departures from the non-spinnning case of $z=0$. The solution valid up to $\mathcal{O}(x^7,z^4)$ can be found in the ancillary Mathematica notebook.

\bibliography{bibliography}

\end{document}